\documentclass[aps,prc,superscriptaddress,showpacs,nofootinbib,floatfix,twocolumn]{revtex4}
\usepackage{graphicx}
\usepackage{dcolumn}
\usepackage{bm}

\newcommand{\snn} {\mbox{$\sqrt{s_{NN}}$}}
\newcommand{\meanpt} {\mbox{$\langle p_{T} \rangle$}}

\newcommand{\pbar} {\mbox{$\overline{p}$}}
\newcommand{\lbar} {\mbox{$\overline{\Lambda}$}}
\newcommand{\npart} {\mbox{$N_{part}$}}

\def\Journal#1#2#3#4{#1 {\bf #2}, #3 (#4)}

\def\IJMPA{Int.\ J.\ Mod.\ Phys.~A}

\def\NIMA{Nucl.\ Instrum.\ Methods~A}
\def\NPA{Nucl.\ Phys.~A}
\def\NPB{Nucl.\ Phys.~B}
\def\PLB{Phys.\ Lett.~B}
\def\PLC{Phys.\ Repts.}
\def\PRL{Phys.\ Rev. Lett.}
\def\PRD{Phys.\ Rev.~D}
\def\PRC{Phys.\ Rev.~C}

\def\EPJC{Eur.\ Phys.\ J.~C}

\begin{document}
\title{Measurement of density correlations in pseudorapidity
via charged particle multiplicity fluctuations in Au+Au collisions at \snn~$=$~200~GeV}

\newcommand{\abilene}{Abilene Christian University, Abilene, TX 79699, USA}
\newcommand{\acadsin}{Institute of Physics, Academia Sinica, Taipei 11529, Taiwan}
\newcommand{\banaras}{Department of Physics, Banaras Hindu University, Varanasi 221005, India}
\newcommand{\barc}{Bhabha Atomic Research Centre, Bombay 400 085, India}
\newcommand{\bnl}{Brookhaven National Laboratory, Upton, NY 11973-5000, USA}
\newcommand{\caucr}{University of California - Riverside, Riverside, CA 92521, USA}
\newcommand{\ciae}{China Institute of Atomic Energy (CIAE), Beijing, People's Republic of China}
\newcommand{\cns}{Center for Nuclear Study, Graduate School of Science, University of Tokyo, 7-3-1 Hongo, Bunkyo, Tokyo 113-0033, Japan}
\newcommand{\columbia}{Columbia University, New York, NY 10027 and Nevis Laboratories, Irvington, NY 10533, USA}
\newcommand{\dapnia}{Dapnia, CEA Saclay, F-91191, Gif-sur-Yvette, France}
\newcommand{\debrecen}{Debrecen University, H-4010 Debrecen, Egyetem t{\'e}r 1, Hungary}
\newcommand{\fsu}{Florida State University, Tallahassee, FL 32306, USA}
\newcommand{\gsu}{Georgia State University, Atlanta, GA 30303, USA}
\newcommand{\hiroshima}{Hiroshima University, Kagamiyama, Higashi-Hiroshima 739-8526, Japan}
\newcommand{\ihepprot}{Institute for High Energy Physics (IHEP), Protvino, Russia}
\newcommand{\isu}{Iowa State University, Ames, IA 50011, USA}
\newcommand{\jinrdubna}{Joint Institute for Nuclear Research, 141980 Dubna, Moscow Region, Russia}
\newcommand{\kaeri}{KAERI, Cyclotron Application Laboratory, Seoul, South Korea}
\newcommand{\kangnung}{Kangnung National University, Kangnung 210-702, South Korea}
\newcommand{\kek}{KEK, High Energy Accelerator Research Organization, Tsukuba-shi, Ibaraki-ken 305-0801, Japan}
\newcommand{\kfki}{KFKI Research Institute for Particle and Nuclear Physics (RMKI), H-1525 Budapest 114, POBox 49, Hungary}
\newcommand{\korea}{Korea University, Seoul, 136-701, Korea}
\newcommand{\kurchatov}{Russian Research Center ``Kurchatov Institute", Moscow, Russia}
\newcommand{\kyoto}{Kyoto University, Kyoto 606-8502, Japan}
\newcommand{\labllr}{Laboratoire Leprince-Ringuet, Ecole Polytechnique, CNRS-IN2P3, Route de Saclay, F-91128, Palaiseau, France}
\newcommand{\lawllnl}{Lawrence Livermore National Laboratory, Livermore, CA 94550, USA}
\newcommand{\losalamos}{Los Alamos National Laboratory, Los Alamos, NM 87545, USA}
\newcommand{\lpc}{LPC, Universit{\'e} Blaise Pascal, CNRS-IN2P3, Clermont-Fd, 63177 Aubiere Cedex, France}
\newcommand{\lund}{Department of Physics, Lund University, Box 118, SE-221 00 Lund, Sweden}
\newcommand{\muenster}{Institut f\"ur Kernphysik, University of Muenster, D-48149 Muenster, Germany}
\newcommand{\myongji}{Myongji University, Yongin, Kyonggido 449-728, Korea}
\newcommand{\nagasaki}{Nagasaki Institute of Applied Science, Nagasaki-shi, Nagasaki 851-0193, Japan}
\newcommand{\newmex}{University of New Mexico, Albuquerque, NM 87131, USA}
\newcommand{\nmsu}{New Mexico State University, Las Cruces, NM 88003, USA}
\newcommand{\ornl}{Oak Ridge National Laboratory, Oak Ridge, TN 37831, USA}
\newcommand{\orsay}{IPN-Orsay, Universite Paris Sud, CNRS-IN2P3, BP1, F-91406, Orsay, France}
\newcommand{\pnpi}{PNPI, Petersburg Nuclear Physics Institute, Gatchina, Russia}
\newcommand{\riken}{RIKEN (The Institute of Physical and Chemical Research), Wako, Saitama 351-0198, JAPAN}
\newcommand{\rikjrbrc}{RIKEN BNL Research Center, Brookhaven National Laboratory, Upton, NY 11973-5000, USA}
\newcommand{\saispbstu}{St. Petersburg State Technical University, St. Petersburg, Russia}
\newcommand{\saopaulo}{Universidade de S{\~a}o Paulo, Instituto de F\'{\i}sica, Caixa Postal 66318, S{\~a}o Paulo CEP05315-970, Brazil}
\newcommand{\seoulnat}{System Electronics Laboratory, Seoul National University, Seoul, South Korea}
\newcommand{\stonybrkc}{Chemistry Department, Stony Brook University, SUNY, Stony Brook, NY 11794-3400, USA}
\newcommand{\stonycrkp}{Department of Physics and Astronomy, Stony Brook University, SUNY, Stony Brook, NY 11794, USA}
\newcommand{\subatech}{SUBATECH (Ecole des Mines de Nantes, CNRS-IN2P3, Universit{\'e} de Nantes) BP 20722 - 44307, Nantes, France}
\newcommand{\tenn}{University of Tennessee, Knoxville, TN 37996, USA}
\newcommand{\titech}{Department of Physics, Tokyo Institute of Technology, Tokyo, 152-8551, Japan}
\newcommand{\tsukuba}{Institute of Physics, University of Tsukuba, Tsukuba, Ibaraki 305, Japan}
\newcommand{\vandy}{Vanderbilt University, Nashville, TN 37235, USA}
\newcommand{\waseda}{Waseda University, Advanced Research Institute for Science and Engineering, 17 Kikui-cho, Shinjuku-ku, Tokyo 162-0044, Japan}
\newcommand{\weizmann}{Weizmann Institute, Rehovot 76100, Israel}
\newcommand{\yonsei}{Yonsei University, IPAP, Seoul 120-749, Korea}
\affiliation{\abilene}
\affiliation{\acadsin}
\affiliation{\banaras}
\affiliation{\barc}
\affiliation{\bnl}
\affiliation{\caucr}
\affiliation{\ciae}
\affiliation{\cns}
\affiliation{\columbia}
\affiliation{\dapnia}
\affiliation{\debrecen}
\affiliation{\fsu}
\affiliation{\gsu}
\affiliation{\hiroshima}
\affiliation{\ihepprot}
\affiliation{\isu}
\affiliation{\jinrdubna}
\affiliation{\kaeri}
\affiliation{\kangnung}
\affiliation{\kek}
\affiliation{\kfki}
\affiliation{\korea}
\affiliation{\kurchatov}
\affiliation{\kyoto}
\affiliation{\labllr}
\affiliation{\lawllnl}
\affiliation{\losalamos}
\affiliation{\lpc}
\affiliation{\lund}
\affiliation{\muenster}
\affiliation{\myongji}
\affiliation{\nagasaki}
\affiliation{\newmex}
\affiliation{\nmsu}
\affiliation{\ornl}
\affiliation{\orsay}
\affiliation{\pnpi}
\affiliation{\riken}
\affiliation{\rikjrbrc}
\affiliation{\saispbstu}
\affiliation{\saopaulo}
\affiliation{\seoulnat}
\affiliation{\stonybrkc}
\affiliation{\stonycrkp}
\affiliation{\subatech}
\affiliation{\tenn}
\affiliation{\titech}
\affiliation{\tsukuba}
\affiliation{\vandy}
\affiliation{\waseda}
\affiliation{\weizmann}
\affiliation{\yonsei}
\author{S.S.~Adler}	\affiliation{\bnl}
\author{S.~Afanasiev}	\affiliation{\jinrdubna}
\author{C.~Aidala}	\affiliation{\bnl}
\author{N.N.~Ajitanand}	\affiliation{\stonybrkc}
\author{Y.~Akiba}	\affiliation{\kek} \affiliation{\riken}
\author{J.~Alexander}	\affiliation{\stonybrkc}
\author{R.~Amirikas}	\affiliation{\fsu}
\author{L.~Aphecetche}	\affiliation{\subatech}
\author{S.H.~Aronson}	\affiliation{\bnl}
\author{R.~Averbeck}	\affiliation{\stonycrkp}
% ...
\collaboration{PHENIX Collaboration} \noaffiliation

\date{\today}

\begin{abstract}
Longitudinal density correlations of produced matter in Au+Au collisions
at \snn~$=$~200~GeV have been measured from the
inclusive charged particle distributions as a function of pseudorapidity window sizes.
The extracted $\alpha \xi$ parameter, 
related to the susceptibility of the density fluctuations in
the long wavelength limit, exhibits a non-monotonic behavior as a function of 
the number of participant nucleons, \npart.
A local maximum is seen at \npart~$\sim$~90, with corresponding energy density
based on the Bjorken picture of $\epsilon_{Bj}\tau$~$\sim$~2.4~GeV/(fm$^{2}\cdot c)$ with
a transverse area size of 60~fm$^{2}$.  This behavior may suggest a critical 
phase boundary based on the Ginzburg-Landau framework.
\end{abstract}


\maketitle
	
\section{INTRODUCTION}
\label{sec:intro}
Theoretical studies of Quantum Chromodynamics~(QCD) in
non-perturbative regimes indicate that QCD matter has a rich
phase structure~\cite{QCDDIAGRAM}.
The phase diagram can be parameterized by
temperature $T$ and baryo-chemical potential $\mu_B$.
Based on the phase diagram, we can obtain perspectives on 
how the vacuum structure of the early universe evolved in
extremely high temperature states after the Big Bang
as well as what happens in extremely high baryon density
states such as in the core of neutron stars.  Therefore,
a comprehensive and quantitative understanding of the QCD phase 
diagram is one of the most important subjects in modern nuclear physics.
At a minimum we expect the phase diagram to exhibit at least two
distinct regions: the deconfined phase where the basic degrees 
of freedom of QCD, quarks and gluons, emerge;
and the hadron phase where quarks and gluons are confined.
There is a first order phase boundary at $\mu_B > 0$ and $T=0$
~\cite{FIRST0, FIRST1, FIRST2, FIRST3, FIRST4, FIRST5, FIRST6, FIRST7}.
At $\mu_B = 0$ and $T>0$ a smooth crossover transition is expected
due to finite masses of quarks~\cite{CROSSOVER}.
Logically we can then expect that a critical end-point (CEP) 
exists at the end of the first order phase transition line~\cite{ENDPOINT}.
The location of the CEP would be a landmark in understanding 
the whole structure of the phase diagram.  Although numerical calculations 
using lattice gauge theory, as well as model calculations, predict the 
existence of the CEP, none of them have reached a quantitative 
agreement on the location at present precision~\cite{QCDDIAGRAM}.
Therefore experimental investigations are indispensable to pin down
the location, and to establish properties of the phase point
based on fundamental observables. 

Strongly interacting, high-density matter has been created in
nucleus-nucleus collisions at Relativistic Heavy Ion Collider~(RHIC) 
at Brookhaven National Laboratory~(BNL)~\cite{WP}.
Strong suppression of hadrons at high transverse momentum~($p_T$)
observed in central Au+Au collisions at $\sqrt{s_{NN}}=200$GeV at RHIC
indicate creation of high density matter~\cite{PPG014, PPG023}.
Strong elliptic flow indicates that the matter thermalizes rapidly 
and behaves like a fluid with very low viscosity~\cite{PPG066}.
Furthermore, the valence quark number scaling
of elliptic flow suggests that quark-like degrees of freedom are 
pertinent in the evolution of the flow~\cite{V2SCALING}.
Those observations naturally lead us to the expectation that the initial 
thermalized state of matter is at $T > T_c$ in central Au+Au collisions, 
and possibly at $T < T_c$ in the most peripheral collisions.
Therefore a system with initial $T=T_c$ may exist somewhere
between peripheral and central collisions.

Since there could be different $T_c$'s depending on order parameters
in the crossover transition~\cite{DIFFTC}, it is worth measuring different
kinds of order parameters.  It is known that density correlations in 
matter are  robust observables for critical temperatures in general~\cite{OZ}.
The order parameter we will focus on here is spatial density fluctuations.
Following the Ginzburg-Landau (GL) framework~\cite{GL}
we expect a correlation 
between fluctuations in density at different points which lead 
to a two-point correlation function of the form of $\alpha e^{-r/\xi}$, 
where $r$ is the one dimensional distance, 
$\alpha$ is the strength of the correlation, and
$\xi \propto |T-T_c|^{-1/2}$ is the spatial correlation length.
This functional form can be derived from the GL
free energy density by expanding it with a scalar order parameter 
which is small enough (see Appendix \ref{ap:deri}).
A large increase of $\xi$ near $T_c$ can be a good indicator 
for a phase transition.
In addition to $\xi$ itself, the product $\alpha \xi$ can also 
be a good indicator of a phase transition.
As shown in Sec.~\ref{sec:observable},
$\alpha \xi$ behaves as $ |1-T_c/T|^{-1}$.
In the GL framework, this quantity is related to the medium's 
susceptibility in the long wavelength limit.
(See Appendix \ref{ap:deri} for the derivation).
The matter produced in the collision 
expands longitudinally from its earliest time, which leads to 
cooling after the initial thermalization.  If the system's 
evolution takes it near a critical point as it cools,
then the large correlated density fluctuations will appear as 
$T$ approaches $T_c$ from above.  If the expansion after that 
point is rapid enough then these fluctuations can potentially
survive into the final state~\cite{CHARGEDIFFUSION}.

Experimentally, spatial density fluctuations in longitudinal
space $z$ in the early stage of an $A+A$ collision at RHIC
can be measured as the density fluctuation in rapidity, or
pseudorapidity, space in the final state.
The differential length $dz$ between neighboring medium elements 
at a common proper time $\tau =\sqrt{t^2-z^2}$
is expressed as $dz=\tau \cosh(y) dy$, where $y$ is rapidity.
If we limit the study to only a narrow region around midrapidity, then
$dz \sim \tau dy$ is valid with the approximation of $\cosh(y)\sim 1$.
Therefore we can observe density fluctuation in $z$ coordinate
as being mapped onto density fluctuations in rapidity space.
In the region around midrapidity used in this analysis
we can approximate rapidity by pseudorapidity ($\eta$) for inclusive charged particles,
since the mean \meanpt~(\meanpt=0.57~GeV/c $\gg m_{\pi}$) observed in
\snn~$=200$GeV collisions at RHIC is so high.

In this paper we measure charged particle density correlations 
in pseudorapidity space to search for the critical phase boundary
in Au+Au collisions at $\sqrt{s_{NN}}=200$GeV.
The density correlation is extracted from inclusive charged particle 
multiplicity distributions measured as a function of pseudorapidity 
window size~$\delta\eta$.
Negative Binomial Distributions~(NBD) are fit to the
measured multiplicity distributions, and the NBD parameters
$\mu$~(mean) and $k^{-1}$~(deviation from a Poissonian width)
are determined.
The product of the correlation strength $\alpha$ and
the correlation length $\xi$ is extracted from a known
relation between the product of $\alpha \xi$ and
the NBD $k$ parameter as a function of $\delta \eta$.
We expect a monotonic correspondence between initial temperature
and measured energy density based on Bjorken picture~\cite{BJ}
which in turn has a monotonic relation with the number of participant
nucleons \npart~in a collision~\cite{PPG019}.
Thus the critical behavior of
$\alpha \xi$ near $T_c$  can be observed as
a non-monotonic increase as a function of $N_{part}$.

It is worth noting that most of experimentally accessible
points on the phase diagram are neither phase boundaries nor
the end-point.  Therefore, before searching for a special phase
point such as CEP based on particular theoretical assumptions,
we would rather observe and define phase boundaries by general 
methods.  The application of the GL framework for density 
correlations far from $T_c$ provides this approach. 
It is known that the GL framework is not applicable 
directly at $T = T_c$ because the fluctuations become too 
large to be described consistently.
The correlation length $\xi$ can not be defined at $T_c$,
where many length scales from small to large emerge.
This is the origin of the power law behavior, or fractal nature 
of fluctuations at the critical phase boundary. 
However, in the regions relatively far from $T_c$,
the fluctuations are naturally expected to be small.
Therefore the GL approach is suitable in the nuclear collision
environment as long as the system approaches a phase boundary from
a thermalized state with $T$ well above $T_c$.  As a future prospect,
once we define a phase boundary even in the crossover region,
we can further investigate the characteristic nature of the phase point,
such as critical exponents based on the chiral condensate
~\cite{FRAME, OPA, DENSITY} along the phase boundary, to judge
whether the point corresponds to CEP or not.

The organization of this paper is as follows.
Sec.~\ref{sec:observable} provides the exact definition of the experimental 
observables mentioned briefly above.
Sec.~\ref{sec:detector} describes the PHENIX detector used to make
the measurements.
Sec.~\ref{sec:analysis} describes the event samples used for this analysis
and the method for corrections applied to the measured multiplicity fluctuations.
The systematic errors on the measured fluctuations are also explained in this section.
In Sec.~\ref{sec:result},
fit results of the NBD parameters in each collision centrality and pseudorapidity
window size are presented, and
the behaviors of the $\alpha\xi$ product as a function of \npart~are presented.
In Sec.~\ref{sec:discussion}, 
in addition to the discussion on the observed \npart~dependence of $\alpha\xi$, 
other possible sources of correlation between inclusive charged particles
are discussed.  The relation between the measured energy density and \npart~is 
discussed to relate \npart~to the initial temperature.
Conclusions are given in Sec.~\ref{sec:conclusion}.
In Appendix~\ref{ap:deri}, the density correlation length and susceptibility are
exactly defined based on the GL framework.
Finally, in Appendix~\ref{ap:table} all measured NBD parameters in all collision
centralities are tabulated.

\section{EXPERIMENTAL OBSERVABLES}
\label{sec:observable}
In this analysis the density fluctuation will be discussed via charged
particle multiplicity distributions as a function of the pseudorapidity
window size for each collision centrality or \npart~range.

It is known that the charged particle multiplicity distributions 
are empirically well described by the Negative Binomial Distribution~(NBD)
in $A+A$, $p+p$ and $e^{+}e^{-}$ collisions~\cite{DREMIN}.
The distribution is expressed as
\begin{eqnarray}
\label{eq:nbd}
P_{k,\mu}(n) = 
\frac{\Gamma(n+k)}{\Gamma(n-1)\Gamma(k)} 
\left( \frac{\mu/k}{1+\mu/k} \right) \frac{1}{1+\mu/k},
\end{eqnarray}
Here $\mu$ is the mean of the distribution and $k^{-1}$ corresponds to the
difference between its width and that of a Poisson with that mean.
Thus the NBD coincides with the Poisson distribution in the case of 
$k=\infty$, and with the Bose-Einstein distribution in the case of $k=1$.
In this sense, the NBD $k$ directly reflects the degree of correlation
between the particles produced into the experimental window.

We can relate the $k$ parameter for the multiplicity distribution
within an $\eta$ window to the correlation between phase-space 
densities in different $\eta$ windows.  Specifically $k$
can be mathematically related with the second order normalized
factorial moment~$F_2$
\begin{eqnarray}
\label{eq:kf2}
k^{-1} = F_2-1
\end{eqnarray}
where $F_2$ corresponds the integrated two-particle correlation function, 
which can be expressed as~\cite{F2}
\begin{eqnarray}
\label{eq:f2c2}
F_2(\delta\eta) &=& \frac{ \langle n(n-1) \rangle }{ \langle n\rangle^2} =
\frac{\int\!\!\int^{\delta\eta}\rho_2(\eta_1,\eta_2) 
d\eta_1 d\eta_2 }{\{\int^{\delta\eta} \rho_1(\eta) d\eta\}^2} \nonumber \\
&=& \frac{1}{(\delta\eta)^2} \int\!\!\int^{\delta\eta} 
\frac{C_2(\eta_1,\eta_2)}{\bar{\rho_1}^2} d\eta_1 d\eta_2 + 1,
\end{eqnarray}
where $n$ is the number of produced particles and $\delta\eta$ is 
the pseudorapidity window size inside which the multiplicities are measured.
In Eq.~(\ref{eq:f2c2}) we introduced one- and two-particle inclusive 
multiplicity densities $\rho_1$ and $\rho_2$ based on the inclusive differential 
cross section relative to the total inelastic cross section $\sigma_{inel}$
as follows~\cite{DREMIN}
\begin{eqnarray}
\label{eq:f2}
\frac{1}{\sigma_{inel}}d\sigma &=& \rho_1(\eta)d\eta, \nonumber \\
\frac{1}{\sigma_{inel}}d^2\sigma &=& \rho_2(\eta_1, \eta_2)d\eta_1 d\eta_2.
\end{eqnarray}
Here $\bar{\rho_1}$ is the average density per unit length within $\delta\eta$
which is defined as
\begin{eqnarray}
\label{eq:rho}
\bar{\rho_1}=\frac{1}{\delta\eta}\int^{\delta \eta} \rho_1(\eta) d\eta.
\end{eqnarray}
With these densities, the two particle density correlation function is defined as
\begin{eqnarray}
\label{eq:c2}
C_2(\eta_1,\eta_2)=\rho_2(\eta_1,\eta_2) - \rho_1(\eta_1)\rho_1(\eta_2).
\end{eqnarray}
Instead of measuring $C_2$ or $F_2$ directly, in this analysis we extract
the NBD $k$ parameter as a measure of particle correlations over $\eta$.
This is partly for historical reasons~\cite{E802}, but also because,
as shown in Sec.~\ref{sec:analysis}, we can correct the measurement
of $k$ for the detector imperfections in a very robust way
by using a statistical property of NBD, while the same correction made
at the level of $F_2$ would require additional information
on the parent distribution.

The normalized two particle correlation function~$C_2$ in the experiment can 
be parametrized as follows, based on the one-dimensional functional form obtained 
in the GL framework (see Eq.~(\ref{eq:fg2})):
\begin{eqnarray}
\label{eq:nomc2}
\frac{C_2(\eta_1, \eta_2)}{\bar{\rho_1}^2} 
= \alpha e^{-|\eta_{1}-\eta_{2}|/ \xi} + \beta,
\end{eqnarray}
where $\bar{\rho_1}$ is proportional to the mean multiplicity in each 
collision centrality bin, or range of \npart, and the scale factor $\alpha$ 
is the strength of the correlations at the zero separation.
The constant term $\beta$ arises from any kind of experimental and physical 
correlations which are independent of the pseudorapidity separation, such
as the residual effect of finite centrality binning.

Further, one has to take into account the fact that the damping behavior 
in Eq.~(\ref{eq:fg2}) is caused only by the spatial inhomogeneity of the
system at a fixed temperature.  In realistic collisions and event
samples there is no single relevant temperature.
For instance, finite centrality binning adds together a range of
fluctuations originating from collisions with different \npart.
However, in principle these centrality-correlated fluctuations are 
independent of the thermally-induced spatial fluctuations. 
In addition, although the self correlation at the zero distance between 
two sub-volumes in Eq.~(\ref{eq:g2}) was excluded, the self correlation 
cannot be excluded in the integrated two particle correlation function contained 
in Eq.~(\ref{eq:f2c2}).  We have tried various kind of functional forms for 
$C_2$ which contained power terms and also plural correlation lengths.  However, 
we found empirically that just adding the constant term in 
Eq.~(\ref{eq:nomc2}) produced the best fit results to all data points.

Finally, the relation between the NBD $k$ parameter and the pseudorapidity
window size $\delta\eta$ can be obtained by the substitution of
Eq.~(\ref{eq:nomc2}) into Eq.~(\ref{eq:f2c2})~\cite{E802, NBDCORR}
\begin{eqnarray}
\label{eq:kintc2}
k^{-1}(\delta\eta) = F_2 - 1 =
\frac{2\alpha \xi^2(\delta\eta/\xi-1+e^{-\delta\eta/\xi})}{\delta\eta^2}+\beta.
\end{eqnarray}
In the limit of $\xi \ll \delta\eta$, which we believe holds in this analysis,
Eq.~(\ref{eq:kintc2}) can be approximated as
\begin{eqnarray}
\label{eq:app}
k(\delta\eta) = \frac{1}{2\alpha\xi/\delta\eta + \beta} & (\xi \ll \delta\eta),
\end{eqnarray}
where experimentally we can not resolve $\alpha$ and $\xi$ separately,
but the product $\alpha\xi$ can be directly determined. The product is 
related to the susceptibility in the long wavelength limit, 
$\chi_{\omega=0} \propto |T-T_c|^{-1}$ for a given temperature
$T$ based on Eq.~(\ref{eq:sus}). 
Combined with the parametrization in Eq.~(\ref{eq:nomc2}),
the $\alpha\xi$ product should then follow
\begin{eqnarray}
\label{eq:alphaxi}
\alpha\xi \propto \bar{\rho_1}^{-2}\frac{1}{|1-T_c/T|}.
\end{eqnarray}
Since we expect that $\bar{\rho_1}$ is a monotonic function of $T$,
in the limit of $T$ far from $T_c$, $\alpha\xi$ should vary monotonically 
as a function of $T$.
However, if $T$ approaches $T_c$, the $\alpha\xi$ product will show 
a singular behavior.  Therefore, any non-monotonic increase of $\alpha\xi$ 
could be an indication of $T \sim T_c$ near a critical point.
If the experimental bias term $\beta$ is excluded in Eq.~(\ref{eq:app}), 
the slope in $k$ versus $\delta\eta$ thus contains crucial information on 
the phase transition.

It is worth mentioning that in this method, correlations on scales even
smaller than the minimum $\delta\eta$ window can be meaningfully discussed
based on the differences of the NBD $k$ as a function of $\delta\eta$ window sizes,
since the correlations are always integrated from the limit of the detector 
resolution to $\delta\eta$ window size.

\section{PHENIX DETECTOR}
\label{sec:detector}
PHENIX is one of four experiments operating at RHIC~\cite{PHENIXNIM}.
The PHENIX detector has two central spectrometer arms, denoted East and West.
Each central arm covers the pseudorapidity range $|\eta|$~$<$~0.35 and subtends
an azimuthal angle range $\Delta\phi$ of $\pi/2$ around the beam 
axis~($z$~direction).
PHENIX includes global detectors which provide information for event triggers
as well as measurement of collision points along the beam axis and collision 
centralities.
A detailed description of the PHENIX detector can be found in~\cite{PHENIXNIM}.
The detector subsystems relevant for this analysis will be briefly 
explained below. 

Charged particles are measured by a drift chamber~(DC) and two multi-wire chambers
with pad readout~(PC1 and PC3) located at 2.2, 2.5 and 5~m from the beam axis in
the East arm, respectively.
The collision vertex points were measured using the time difference between
two Beam-Beam Counters~(BBC) located at z~$=$~+144~cm~(north side) and
z~$=$~-144~cm~(south side) from the nominal interaction point (IP) along the
beam line, which cover pseudorapidity ranges of 3.0~$<$~$\eta$~$<$~3.9~(north)
and -3.9~$<$~$\eta$~$<$~-3.0~(south), respectively.
Each BBC has 64~$\check{C}$erenkov counter elements with the typical time
resolution of 50~ps.  Combined with BBC's, two Zero Degree Calorimeters~(ZDC)
were further used. The ZDC's are designed to measure energies of spectator neutrons
within a cone of 2~mrad around the beam axis.
The two ZDC's are located at z~$=$~$\pm$~18.25~m from IP, respectively.
The Au+Au minimum bias trigger and collision centralities were provided
by combining information from both BBC's and ZDC's.

\section{DATA ANALYSIS}
\label{sec:analysis}
\subsection{Run and Event selection}
We have used data taken in Au+Au collisions at $\snn$~$=$~200~GeV with the magnetic
field off condition during RHIC Run-2 in~2002, in order to optimize acceptance for
the low $p_T$ charged particles.
The basic trigger required coincident hits in the two BBC's
(equal or more than two hit $\check{C}$erenkov elements in each side) and the two
ZDC's (equal or more than one neutron in each side).
The efficiency of this minimum-bias trigger is estimated as 92.2$^{+2.5}_{-3.0}$\%
to the total Au+Au inelastic cross section by the Monte Carlo (MC) simulation based 
on the Glauber model~\cite{PPG014}.
Events with collision points within $\pm$~5~cm from the nominal IP as measured by the BBC
were analyzed.
In total, 258k events taken by the minimum-bias trigger were used in this analysis. 
We have rigorously checked the detector stability by looking at multiplicity
correlations between the relevant sub-detector systems, as well as by monitoring
positions of inefficient areas over the span of the analyzed dataset.
We allowed 2\% fluctuation on the average multiplcity of measured number
of charged tracks in entire analyzed run ranges.

\subsection {Track selection}
\label{subsec:track}
\begin{figure}[tbh]
\includegraphics[width=1.0\linewidth]{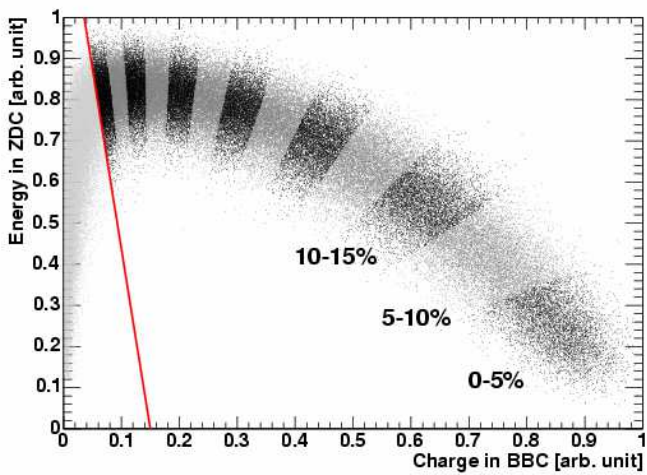}
\caption{
(Color online)
Definition of collision centrality, BBC charges versus ZDC energy.
Event samples in 5\% bin width are plotted from 0~-~5\%~(central)
to 60~-~65\%~(peripheral).
The solid line indicates the limit of the most peripheral sample used
for this analysis.
}
\label{fig:centdef}
\end{figure}
In this analysis, charged tracks detected in the East arm
($|\eta|$~$<$~0.35, $\Delta\phi$~$<$~$\pi/2$) were used.
As charged track selection criteria, we required that each straight-line 
track reconstructed by a DC hit pattern associated with a PC1 hit
be aligned with a PC3 hit and the collision vertex point measured by BBC.
We required associations between DC tracks and PC3 hits to be
within 10~cm in the distance of closest approach~(DCA),
which was determined to minimize the random associations.
The DC has six types of wire modules; two of them are used for the track 
reconstruction for the azimuthal angle and others are used for the pattern 
recognition.
Selected tracks were reconstructed by using all wire modules of DC.

In addition to the single track selection, we required a minimum
two-track separation in order to minimize effects from fake tracks 
and associated secondary particles.
When we find tracks within the minimum separation window of $\delta\eta$~$<$~0.001
and $\delta\phi$~$<$~0.012~rad, we count them as one track independent of the
number of reconstructed tracks in the window.
These cut values were determined by looking at $\delta\eta$ and $\delta\phi$ 
distributions on the $\eta-\phi$ plane of any two track pairs in the real data sample.
The DC track resolution of 2~mm in the $z$ direction at a reference radius 
of 220~cm from the beam axis corresponds to 1.0~$\times$~$10^{-3}$ in $\eta$.
PC1 and PC3, which are used for the track association, have the same solid angle
each other, and these pixel sizes are 8.4~mm and 14.7~mm, respectively.
These pixel sizes are greater than the requirement of two-track separation cuts,
however, these resolutions are 1.7~mm and 3.6~mm for PC1 and PC3 respectively in
z direction, and these values also corresponds to 1.0~$\times$~$10^{-3}$ in $\eta$.
The resolution in $\phi$ is 1~mrad, but the maximum drift length in DC 
corresponds to 0.012~rad. Therefore the two-track separation window size in
$\eta$ and $\phi$ is consistent with what is expected.

In the case of normal magnetic field condition at the PHENIX detector, which
is used to identify the charged particles, the threshold transverse
momenta $p_T$ correspond to 0.2~GeV/$c$, 0.4~GeV/$c$ and 0.6~GeV/$c$ for
charged pions $\pi^{\pm}$, charged kaons $K^{\pm}$ and protons $p$(antiprotons $\bar{p}$),
respectively~\cite{PPG026}.
Since this analysis used the data taken without magnetic field,
the threshold transverse momenta $p_T$ can be lowered to 0.1~GeV/$c$, 0.25~GeV/$c$ and
0.35~GeV/$c$ for $\pi^{\pm}$, $K^{\pm}$ and $p$($\bar{p}$), respectively.
They were estimated by the GEANT-based Monte Carlo (MC)~\cite{GEANT} simulation
by requiring the equivalent single track selection criteria.
The average transverse momentum $p_T$ for the detected inclusive charged particles
used in this analysis corresponds to 0.57~GeV/$c$, which was also estimated by 
using the measured $p_T$ spectra~\cite{PPG026} with the MC simulation.
Therefore, the difference of the rapidity and pseudorapidity is
at most 3\% at the edge of the PHENIX acceptance.

\subsection {Centrality definition and the number of participant nucleons \npart}
\label{subsec:cent}
The collision centrality was determined by looking the correlation between
a deposited charge sum in both north and south BBC's and an energy sum in 
both ZDC's on an event-by-event basis.  As shown in Fig.~\ref{fig:centdef}, 
the centrality percentile is defined as the fraction of the number of
events in a selected centrality bin on the correlation plot to 
the total number of minimum bias events, corrected for the
min-bias trigger efficiency.
Each axis is normalized to its maximum dynamic range.
As the standard centrality definition, we adopt 5\% centrality bin width
from 0~-~5\%(central) to 60~-~65\%(peripheral) as indicated in the figure.
The lower limit of 65\% is indicated by the solid line in the figure.
In the following analysis, as control samples, we also adopt 10\% bin width 
by merging two 5\% bin width samples from 0~-~10\% to 50~-~60\% and 
from 5~-~15\% to 55~-~65\%. The latter is referred to as a 5\% shifted 10\% bin width.
It is worth noting that the change of the centrality
bin width shifts the mean values in the charged particle multiplicity
distributions, which becomes a strict systematic check on parameter extractions
with different event ensembles, even with the same total event sample.

Mapping the centralities to the number of participant nucleons, \npart, is based
on the Glauber model, which is described in detail in~\cite{PPG019}.
The quoted mean \npart~and its error can be obtained from~\cite{PPG026}.
In only the 5\% shifted 10\% bin width case, the mean \npart~and its error
were evaluated by averaging two 5\% centrality bins
and estimated from its error propagations, respectively.

\subsection{Measurement of multiplicity distributions of charged particles}
\label{subsec:measure}

\begin{figure}[tbh]
\includegraphics[width=0.8\linewidth]{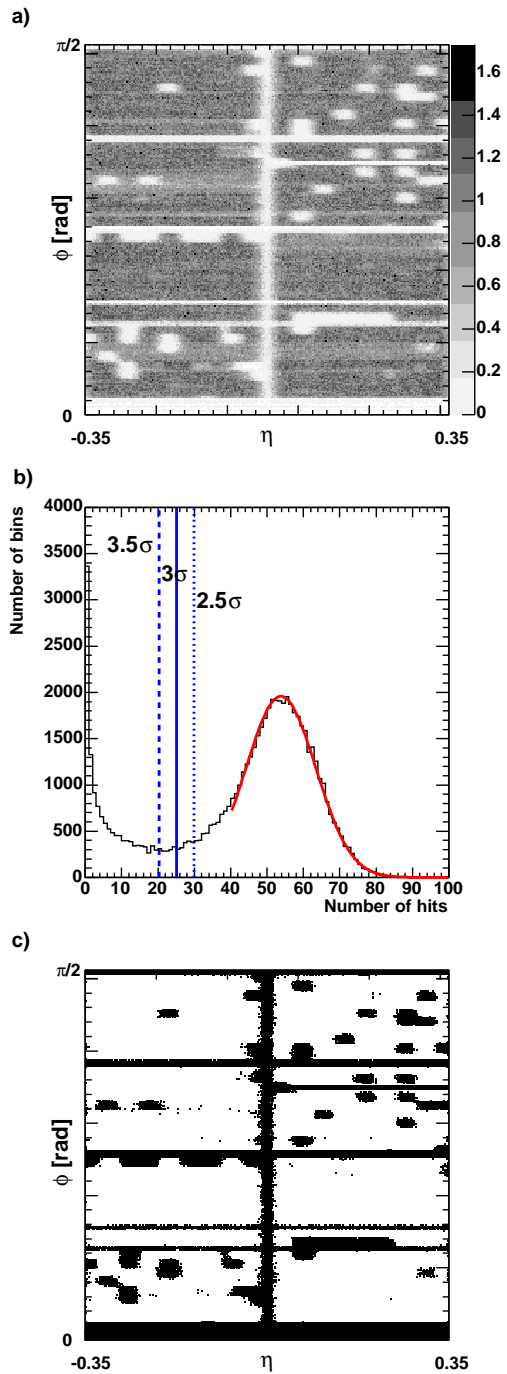}
\caption{
(Color online)
2-dimensional dead map definitions.
a)~Track projection points onto the $\eta-\phi$ plane in the data after
all track selections.
The scale is normalized to the mean number of hits in the peak position in b).
b)~The number of bins among subdivided $2^{8}$~$\times$~$2^{8}$ bins
as a function of the accumulated number of hits over the total event sample.
c)~Definition of the central dead map by excluding the detector region below 3~$\sigma$,
where black parts are identified as dead areas. 
}
\label{fig:deaddef}
\end{figure}

Multiplicity distributions of charged particles were measured
while changing the pseudorapidity window size $\delta\eta$ from 0.066 to 0.7
with a step size of $0.7/2^5$~$=$~0.022. For a given pseudorapidity window size,
the window position in the pseudorapidity axis was shifted by a step of
$0.7/2^{8}$~$=$~0.0027 as long as the window is fully contained within the
PHENIX acceptance of $|\eta|$~$<$~0.35.
For each window position NBD fits were performed to the multiplicity distributions.
Biases originating from inefficient detector areas were corrected with the procedure
explained in Sec.~\ref{subsec:correction}. 
Since even corrected NBD $k$ parameters are not necessarily equal in the case of
extremely inefficient window positions, we have truncated window positions where
the reconstruction efficiency is below 50\%. This truncation is mainly to exclude
biases from the largest hole in the middle of the charged particle detector
as shown in Fig.~\ref{fig:deaddef}~(a) and~(c).
After the truncation, we obtained weighted mean of corrected NBD parameters
$(\langle \mu_c \rangle,$~$\langle k_c \rangle)$ for a given window size, which
are defined as 
\begin{eqnarray}
\label{eq:wmean}
\langle \mu_c \rangle &\equiv&
\sum^n_{i=1} \delta{\mu_c}_i^{-2}{\mu_c}_i / \sum^n_{i=1} \delta{\mu_c}_i^{-2}, \nonumber \\
\langle k_c \rangle &\equiv&
\sum^n_{i=1} \delta{k_c}_i^{-2}{k_c}_i / \sum^n_{i=1} \delta{k_c}_i^{-2},
\end{eqnarray}
where $n$ is the number of valid window positions after the truncation and $\delta$
indicates errors on fitting parameters by the Minuit program~\cite{MINUIT} in each
window position~$i$.
We have performed this procedure in each centrality bin with 5\% and 10\% centrality
bin width, respectively.

The lower limit of 0.066 was determined so that small window sizes, where 
corrected NBD $k$ was seen to depend heavily on window position,
are all excluded.  The lower limit is common for all centrality bins.  

\clearpage

\begin{figure*}[bth]
\includegraphics[width=0.75\linewidth]{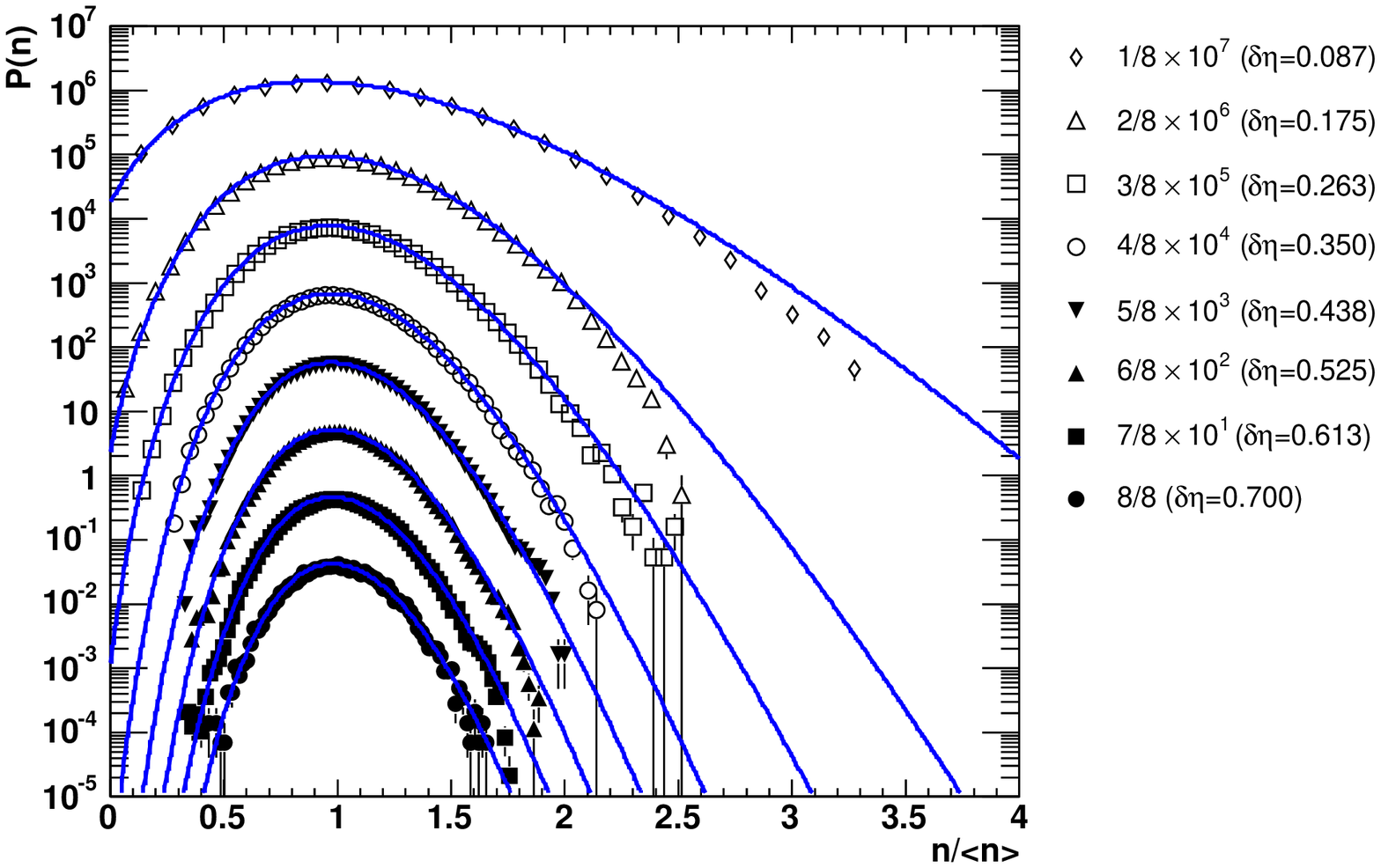}
\caption{
(Color online)
Uncorrected charged particle multiplicity distributions 
in each pseudorapidity window
size, as indicated in the legend, at 0~-~10\% collision centrality.
The distributions are shown as a function of the number of
tracks $n$ normalized to the mean multiplicity $\langle n \rangle$ in each window.
The error bars show the statistical errors.
The solid curves are fit results of NBD.
}
\label{fig:multi}
\end{figure*}

\begin{figure*}[tbh]
\includegraphics[width=1.0\linewidth]{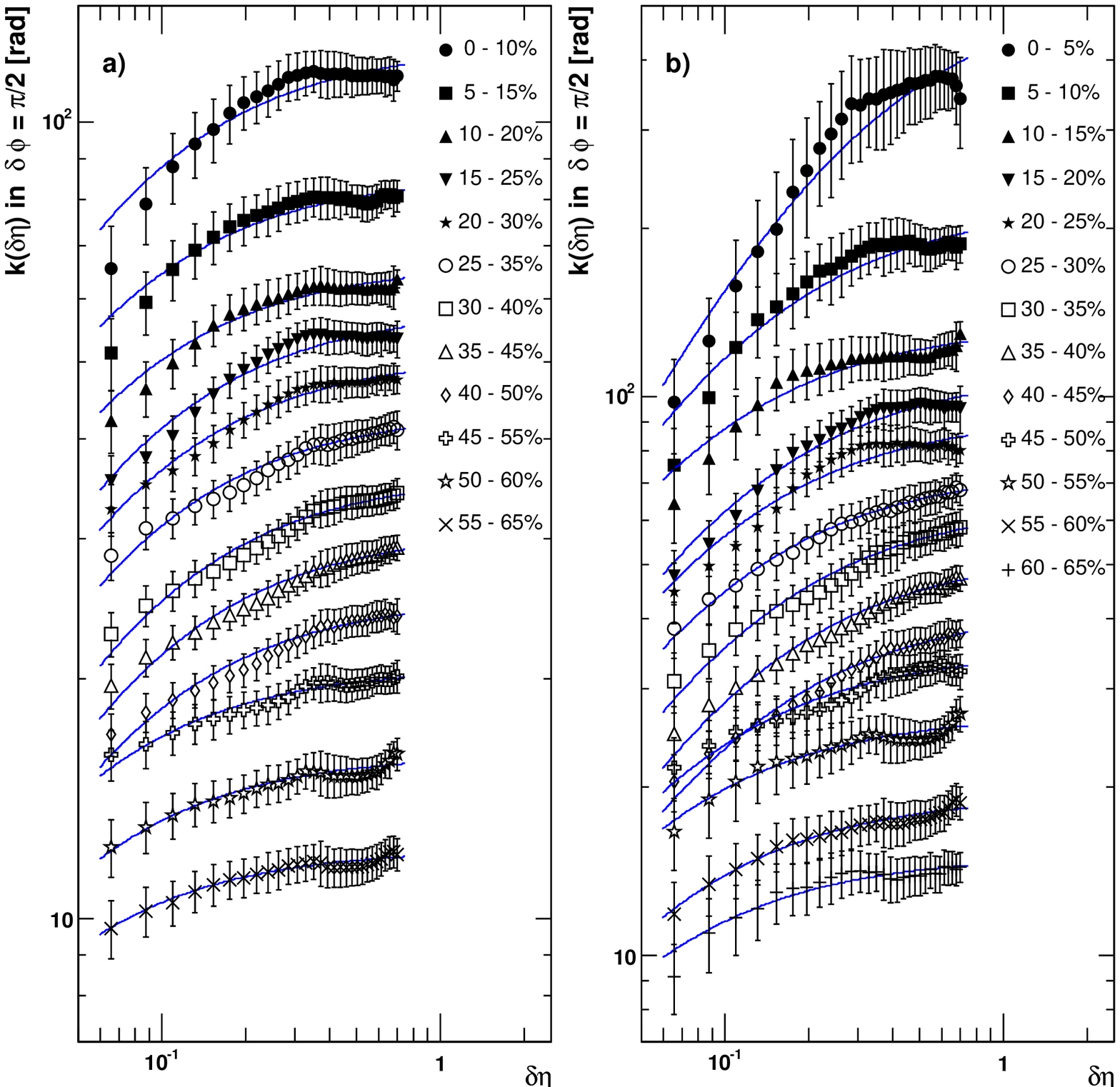}
\caption{\label{fig:kdetachi}  (Color online)
Weighted mean of corrected NBD $k$, $\langle k_c \rangle$ as a function of
pseudorapidity window size with a)~10\% and b)~5\% centrality bin widths.
Centrality classes are indicated in the figure legend.
The error bars show $\delta \langle k_c \rangle$~(total), as
explained in Sec.~\ref{subsec:errors}.
The solid lines indicate the fit curves of Eq.~(\ref{eq:app}).
}
\end{figure*}

\subsection{Correction of NBD $k$ and $\mu$}
\label{subsec:correction}

Any dead or inefficient areas in the detector have been identified and
the bias on the NBD parameters has been corrected based on 
a suitable statistical property of NBD.
Maps of dead areas were produced from the track projection points
onto the $\eta-\phi$ plane in the data after the track selections,
as shown in Fig.~\ref{fig:deaddef}~a), where the detector acceptance is divided into
$2^{8}$~$\times$~$2^{8}$ bins in the $\eta-\phi$ plane.
The accumulated number of hits over the total event sample in each bin is shown
by a gray scale reflecting the statistical weights.
The scale is normalized to the mean number of hits in the peak position shown in
Fig.~\ref{fig:deaddef}~b).
Figure~\ref{fig:deaddef}~b) shows the number of bins among subdivided $2^{8}$~$\times$~$2^{8}$ bins
as a function of the accumulated number of hits over the total event sample
in each $1/2^{8}$~$\times$~$1/2^{8}$ acceptance.
If there were no dead or inefficient area,
a binomial distribution is expected with a probability of $1/2^{8}$~$\times$~$1/2^{8}$
to the total acceptance. For the binomial part, we took a $\pm$~3~$\sigma$ region.
On the other hand, if there are any dead or inefficient areas they tend to
contaminate the lower tail of the binomial distribution.
We defined a central dead map by excluding detector region below 3~$\sigma$ 
as shown in Fig.~\ref{fig:deaddef}~c) where black indicates regions that are
identified as dead areas.
The fraction of good area corresponds to 78\% of the total acceptance.
This map was used to completely suppress particles which hit the dead areas in the
real data.

As long as the baseline distribution is approximated as an NBD, which is certainly
true as observed in E802~\cite{E802} and in the present analysis, one can estimate
the relation between true $k$ values of the NBD and biased $k$ values due to dead
or inefficient areas based on the convolution theorem of NBD.
For two independent NBD's with $(\mu_1, k_1)$ and $(\mu_2, k_2)$,
it is known that the convolution of the two NBD's is an NBD with
$(\mu_c, k_c)$, which satisfies relations as
\begin{eqnarray}
\label{eq:kconv}
k_c &=& k_1 + k_2, \nonumber \\
\mu_c &=& \mu_1 / k_1(k_1+k_2),
\end{eqnarray}
where $\mu_1/k_1$~$=$~$\mu_2/k_2$ holds~\cite{CONV, MJT}.
Therefore the correction can be applied by multiplying a ratio of the total number of
$\eta-\phi$ bins in a given $\eta$ window size to the number of bins excluding dead area,
as the geometrical acceptance corrections can be applied.

Strictly speaking we can not completely reproduce the original $k$ by this correction,
since NBD's in different positions are not completely independent.
However, except for the large hole which is already excluded by the truncation,
small holes are scattered rather uniformly in azimuthal direction for any position
of the $\delta\eta$ windows.  As the simplest overall correction to each window position,
we applied the convolution theorem~\cite{CONV, MJT}
by assuming collection of independent NBD sources.
As long as the correction is applied in the same manner for all the azimuthal holes, 
it does not greatly affect the differential measurement to the pseudorapidity space. 
If the correction is accurate enough, we can expect a constancy of the corrected $k$ values
which should be independent of the fraction of dead areas.
Based on the degree of constancy of corrected $k$ as a function of
the fraction of dead areas in each window position for a given $\delta\eta$ window size,
the incompleteness of the correction in each window size has been checked.
As briefly mentioned in the last paragraph of Sec.~\ref{subsec:measure},
the window sizes to be analyzed were determined so that systematic error bands on
$\langle k_c \rangle$ explained in Sec.~\ref{subsec:errors}, can contain the most of the 
corrected $k$ values independently of the fraction of dead areas in each window position.


\subsection{Statistical and systematic errors}
\label{subsec:errors}
As a convolution of statistical errors, we adopted errors on weighted
mean values $(\delta\langle\mu_c\rangle,$~$\delta\langle k_c\rangle)$ on corrected NBD
parameters after the window truncation mentioned in Sec.~\ref{subsec:measure},
which are defined as 
\begin{eqnarray}
\label{eq:kerr}
\delta\langle\mu_c\rangle^2 &\equiv& \frac{\bar{\delta{\mu_c}_i}^2}{n_{ind}}, \nonumber \\
\delta\langle k_c\rangle^2 &\equiv& \frac{\bar{\delta{k_c}_i}^2}{n_{ind}},
\end{eqnarray}
where $\bar{\delta{\mu_c}_i}$ and $\bar{\delta{k_c}_i}$ are respectively
defined as $\sum^{n}_{i=1}\delta\mu_c/n$ and $\sum^{n}_{i=1}\delta k_c/n$ 
with the number of valid window positions $n$
after the truncation and $n_{ind}\equiv 0.75/\delta\eta$ is the number of statistically
independent window positions for a given $\delta\eta$ window size.
This statistical error on $\delta\langle k_c\rangle$ is referred to as
$\delta\langle k_c\rangle$~(stat).

The dominant sources of systematic errors for the correlation length measurement
are the correction procedure with dead maps and the two-track separation cuts,
since both introduce unphysical correlations.
We have allowed 2\% fluctuation on the average multiplicity of measured number
of charged tracks.
This fluctuation is also 
a result of 
dead channels in the tracking detectors discussed in Sec.~\ref{subsec:track}. 
In order to estimate this, we defined two more patterns of dead maps with the
definition of 3~$\sigma$~$\pm$~0.5~$\sigma$ as indicated in Fig.~\ref{fig:deaddef}~c).
The deviation of $\langle k_c\rangle$ from the central dead map
definition is referred to as $\delta\langle k_c\rangle$~(dead), which
corresponds to 3.4\% typically.

The two-track separation cut serves mainly to reject fake track effects;
these are dominantly observed in the $\phi$ direction rather than $\eta$, since the PC1
hit requirement fixes z positions along the beam axis.
Therefore, the effect of the $\delta\phi$ cut was estimated as $\pm$~0.002~rad
around the central cut value of 0.012~rad with a fixed cut value on $\delta\eta$ of 0.001.
The deviation of $\langle k_c\rangle$ from the central 
value due to the fake track rejection cut 
is referred to as $\delta\langle k_c\rangle$~(fake).
This systematic error increases at higher centrality bins, 
and is estimated as
5.8\% and 0.3\% at 0~-~5\% and 60~-~65\% centrality bins, respectively.


The $\langle k_c \rangle$~(stat) is related to agreement between 
multiplicity distributions and NBD.
The $\langle k_c \rangle$~(dead) and $\langle k_c \rangle$~(fake)
depends on the position of the window and 
the average multiplicity
in a selected centrality bin, respectively.
By treating these contributions as independent systematic error sources,
the total systematic error $\delta \langle k_c \rangle$~(total)
on $\langle k_c \rangle$ in each $\delta\eta$ in each centrality, was obtained
by the quadratic sum over $\delta\langle k_c \rangle$~(stat),
$\delta\langle k_c \rangle$~(dead) and $\delta\langle k_c \rangle$~(fake).

\section{RESULTS}
\label{sec:result}
In this section the results of the NBD fits are first tabulated.
Then the measured NBD $k$ as a function of the pseudorapidity
window sizes in various centrality bins are shown.
Lastly, the \npart~dependences of extracted 
$\alpha\xi$ product in Eq.~(\ref{eq:app}) are presented.

\subsection{NBD fit}

NBD fit results in all window sizes in all centrality bins are summarized
in Appendix Table~\ref{ap_01} through Table~\ref{ap_25} where $\langle \mu_c \rangle$
and $\langle \mu \rangle$ are weighted means of corrected and uncorrected $\mu$
over all window positions respectively, $\langle k_c \rangle$ and $\langle k \rangle$
are weighted means of corrected and uncorrected $k$ over all window positions, respectively.
The $\langle \mu_c \rangle$'s are corrected only for the effect of the detector dead
areas as described in Sec.~\ref{subsec:correction}.
The mean multiplicities were confirmed to be consistent with the result of the independent
analysis by the different method using only PC1 and PC3~\cite{PPG019}, 
after known additional correction factors were taken into account.
Statistical errors on weighted means $\delta \langle k_c \rangle$~(stat) are obtained
as explained in Sec.~\ref{subsec:errors}.
$\langle \chi^2/NDF \rangle$ is the average of reduced $\chi^2$ of NBD fits over all
window positions. $\langle NDF \rangle$ is the average of the degree of freedom of
NBD fits over all window positions, and the systematic errors $\delta \langle k_c \rangle$~(dead),
$\delta \langle k_c \rangle$~(fake) and $\delta \langle k_c \rangle$~(total) are already explained
in Sec.~\ref{subsec:errors}.

The mean and r.m.s. of the reduced $\chi^2$ values in the NBD fit
over all window positions and all $\delta\eta$ sizes and all centralities were obtained as
0.75 and 0.33 respectively. The mean value corresponds to typically 80\%
confidence level. Therefore, it is good enough to assume NBD
as a baseline multiplicity distribution to obtain the integrated
correlation function via the $k$ parameter.

As a demonstration to show how well the NBD fits work,
Figure~\ref{fig:multi} shows the charged particle multiplicity distributions
in each pseudorapidity window size in 1/8 fractions of the full rapidity
coverage of $|\eta|$~$<$~0.35 with 0~-~10\% events in the collision centrality,
where the uncorrected multiplicity distributions within the total error bands
on $\langle k_c \rangle$ in Appendix Table~\ref{ap_01} are all merged.
The distributions are shown as a function of the number of
tracks $n$ normalized to the mean multiplicity $\langle n \rangle$ in each window.
The error bars show the statistical errors on the merged distributions.
The solid curves are fit results with NBD only for the demonstration purpose.
The fit results in Appendix Table~\ref{ap_01} through Table~\ref{ap_25} are
not obtained from these convoluted distributions whose accuracies are 
degraded by the convolutions with different $\mu$ values due to different
detector biases depending on the window positions.

\subsection{$k$ versus $\delta\eta$}

Figures~\ref{fig:kdetachi}~a) and b) show $\langle k_c \rangle$
as a function of pseudorapidity window size with 10\% and 5\% centrality bin
width, respectively.
Centrality classes are indicated inside the figures.
The error bars show $\delta \langle k_c \rangle$~(total) defined in Sec.~\ref{subsec:errors}.
The solid lines in Fig.~\ref{fig:kdetachi} indicate the fit results based on Eq.~(\ref{eq:app}).
The fits were performed in the $\delta\eta$ region from 0.066 to 0.7 as explained in
Sec.~\ref{subsec:measure}.

If we could reliably measure the NBD $k$ parameter for arbitrarily small
$\delta\eta$~$\sim$~0 windows, then $\alpha$ and $\xi$ could be treated
as independent free parameters for each centrality.
In the real experimental situation, there is an anti-correlation  
between $\alpha$ and $\xi$ due to the lack of reliable data points close 
to $\delta\eta$~$\sim$~0~, if we attempt to fit with Eq.(\ref{eq:kintc2}).
However, at least an upper limit on the absolute scale of $\xi$ was obtained
as $\xi$~$<$~0.035 by the free parameter fits based on Eq.~(\ref{eq:kintc2}).
It is qualitatively consistent with expectation 
from numerical calculations~\cite{XIHION}
that the correlation lengths become smaller 
in the RHIC energy than for $p+p$ collisions~\cite{DREMIN} and 
low energy $A+A$ collisions~\cite{E802}.


Since the upper limit of $\xi$ is small enough compared to the fitting region of
$\delta\eta$ ($\xi \ll \delta\eta$), Eq.~(\ref{eq:app}) can be applied for the
fits to the NBD $k$ as a function of $\delta\eta$.
In this case, the $\alpha\xi$ product, which is related to the susceptibility
in the long wavelength limit as defined in Eq.(\ref{eq:sus}), can be obtained by the
fits without any physical assumptions.
The typical $\chi^2/NDF$ in the fit based on Eq.~(\ref{eq:app}) is 0.132, which
corresponds to 99\% confidence level.
Therefore, the small correlation length is confirmed as below the minimum $\delta\eta$
window sizes of 0.066.

As explained in Sec.~\ref{sec:observable} for Eq.~(\ref{eq:app}),
in the limit of $\beta=0$, the slopes in $k$ versus $\delta\eta$
have crucial information on the phase transition. In Fig.~\ref{fig:kdetachi} 
we can identify different behaviors in slopes around 40-50\% centrality region
even without fit curves.


\subsection{$\alpha\xi$ product versus \npart}

\begin{figure}[tbh]
\includegraphics[width=1.0\linewidth]{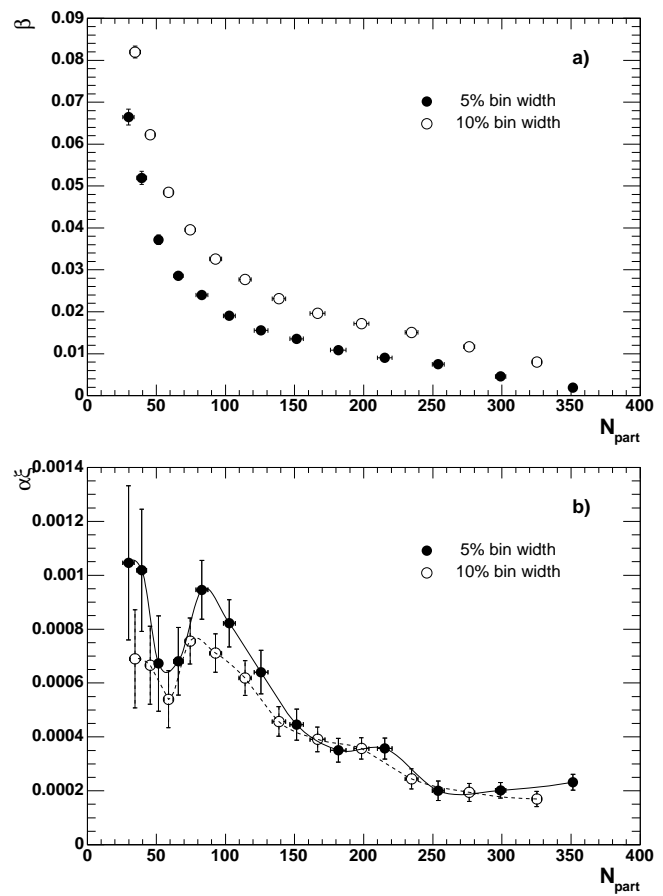}
\caption{
Fit results based on Eq.~(\ref{eq:app}).
a) is $\beta$ and b) is products of $\alpha\xi$ as a function of \npart.
The horizontal error bars correspond to ambiguities in the mean value of~\npart~as
explained in Sec.~\ref{subsec:cent}. The vertical error bars are obtained from errors
on the fitting parameter.
}
\label{fig:bechi}
\end{figure}

Figures~\ref{fig:bechi}~a) and b) show the obtained fit parameters
$\beta$ and $\alpha\xi$ with Eq.~(\ref{eq:app}) as a function of \npart,
where results for both the 5\% and 10\% centrality bin width cases are plotted
as filled and open circles, respectively.
The smooth solid and dotted curves are provided to guide the eye.
The horizontal error bars correspond to ambiguities on the mean value of \npart~as explained
in Sec.~\ref{subsec:cent}.
The vertical error bars are obtained from errors on the fitting parameter by the Minuit
program~\cite{MINUIT}.

Table~\ref{tab:chi} summarizes the fit results where centralities, corresponding \npart,
$\alpha\xi$, $\beta$ and $\chi^2/NDF$ obtained by the fit with Eq.~(\ref{eq:app})
are shown for 10\% and 5\% centrality bin cases respectively.

\begin{table*}[tbh]
\caption{
The $\alpha\xi$ and $\beta$ in Eq.~(\ref{eq:app}) obtained by the fits to
$\langle k_c \rangle$ versus\ $\delta\eta$.
Upper and lower column corresponds to 10\% and 5\% centrality bin width cases, respectively.
}
\footnotesize
\begin{ruledtabular}
\begin{tabular}{ccccc}
Centrality (\%) & $\langle N_{part} \rangle$ & $\alpha\xi~(\propto\chi_{\omega=0})$ & $\beta$ & $\chi^{2}/NDF~(NDF = 27)$ \\
\hline
 $0-10$  & $325.2 \pm 3.3$ & $0.17 \times 10^{-3} \pm 0.03 \times 10^{-3}$ & $0.80 \times 10^{-2} \pm 0.02 \times 10^{-2}$ & $0.24$ \\
 $5-15$  & $276.4 \pm 4.0$ & $0.19 \times 10^{-3} \pm 0.03 \times 10^{-3}$ & $1.17 \times 10^{-2} \pm 0.02 \times 10^{-2}$ & $0.16$ \\
 $10-20$ & $234.6 \pm 4.7$ & $0.24 \times 10^{-3} \pm 0.04 \times 10^{-3}$ & $1.51 \times 10^{-2} \pm 0.03 \times 10^{-2}$ & $0.14$ \\
 $15-25$ & $198.4 \pm 5.4$ & $0.36 \times 10^{-3} \pm 0.04 \times 10^{-3}$ & $1.72 \times 10^{-2} \pm 0.03 \times 10^{-2}$ & $0.26$ \\
 $20-30$ & $166.6 \pm 5.4$ & $0.39 \times 10^{-3} \pm 0.05 \times 10^{-3}$ & $1.96 \times 10^{-2} \pm 0.03 \times 10^{-2}$ & $0.09$ \\
 $25-35$ & $138.6 \pm 4.9$ & $0.46 \times 10^{-3} \pm 0.06 \times 10^{-3}$ & $2.31 \times 10^{-2} \pm 0.04 \times 10^{-2}$ & $0.09$ \\
 $30-40$ & $114.2 \pm 4.4$ & $0.62 \times 10^{-3} \pm 0.06 \times 10^{-3}$ & $2.77 \times 10^{-2} \pm 0.05 \times 10^{-2}$ & $0.13$ \\
 $35-45$ & $92.8  \pm 4.3$ & $0.71 \times 10^{-3} \pm 0.07 \times 10^{-3}$ & $3.26 \times 10^{-2} \pm 0.05 \times 10^{-2}$ & $0.14$ \\
 $40-50$ & $74.4  \pm 3.8$ & $0.76 \times 10^{-3} \pm 0.09 \times 10^{-3}$ & $3.96 \times 10^{-2} \pm 0.07 \times 10^{-2}$ & $0.14$ \\
 $45-55$ & $58.8  \pm 3.3$ & $0.54 \times 10^{-3} \pm 0.11 \times 10^{-3}$ & $4.85 \times 10^{-2} \pm 0.08 \times 10^{-2}$ & $0.05$ \\
 $50-60$ & $45.5  \pm 3.3$ & $0.67 \times 10^{-3} \pm 0.14 \times 10^{-3}$ & $6.22 \times 10^{-2} \pm 0.11 \times 10^{-2}$ & $0.11$ \\
 $55-65$ & $34.6  \pm 3.8$ & $0.69 \times 10^{-3} \pm 0.18 \times 10^{-3}$ & $8.19 \times 10^{-2} \pm 0.14 \times 10^{-2}$ & $0.05$ \\
\hline
 $0-5$   & $351.4 \pm 2.9$ & $0.23 \times 10^{-3} \pm 0.03 \times 10^{-3}$ & $0.19 \times 10^{-2} \pm 0.02 \times 10^{-2}$ & $0.18$ \\
 $5-10$  & $299.0 \pm 3.8$ & $0.20 \times 10^{-3} \pm 0.03 \times 10^{-3}$ & $0.46 \times 10^{-2} \pm 0.02 \times 10^{-2}$ & $0.27$ \\
 $10-15$ & $253.9 \pm 4.3$ & $0.20 \times 10^{-3} \pm 0.04 \times 10^{-3}$ & $0.75 \times 10^{-2} \pm 0.02 \times 10^{-2}$ & $0.17$ \\
 $15-20$ & $215.3 \pm 5.3$ & $0.36 \times 10^{-3} \pm 0.04 \times 10^{-3}$ & $0.90 \times 10^{-2} \pm 0.03 \times 10^{-2}$ & $0.18$ \\
 $20-25$ & $181.6 \pm 5.6$ & $0.35 \times 10^{-3} \pm 0.04 \times 10^{-3}$ & $1.08 \times 10^{-2} \pm 0.03 \times 10^{-2}$ & $0.32$ \\
 $25-30$ & $151.5 \pm 4.9$ & $0.45 \times 10^{-3} \pm 0.06 \times 10^{-3}$ & $1.35 \times 10^{-2} \pm 0.04 \times 10^{-2}$ & $0.02$ \\
 $30-35$ & $125.7 \pm 4.9$ & $0.64 \times 10^{-3} \pm 0.08 \times 10^{-3}$ & $1.55 \times 10^{-2} \pm 0.05 \times 10^{-2}$ & $0.09$ \\
 $35-40$ & $102.7 \pm 4.3$ & $0.82 \times 10^{-3} \pm 0.09 \times 10^{-3}$ & $1.90 \times 10^{-2} \pm 0.05 \times 10^{-2}$ & $0.08$ \\
 $40-45$ & $82.9  \pm 4.3$ & $0.95 \times 10^{-3} \pm 0.11 \times 10^{-3}$ & $2.40 \times 10^{-2} \pm 0.07 \times 10^{-2}$ & $0.06$ \\
 $45-50$ & $65.9  \pm 3.4$ & $0.68 \times 10^{-3} \pm 0.13 \times 10^{-3}$ & $2.86 \times 10^{-2} \pm 0.08 \times 10^{-2}$ & $0.08$ \\
 $50-55$ & $51.6  \pm 3.2$ & $0.67 \times 10^{-3} \pm 0.18 \times 10^{-3}$ & $3.72 \times 10^{-2} \pm 0.11 \times 10^{-2}$ & $0.11$ \\
 $55-60$ & $39.4  \pm 3.5$ & $1.02 \times 10^{-3} \pm 0.23 \times 10^{-3}$ & $5.19 \times 10^{-2} \pm 0.16 \times 10^{-2}$ & $0.06$ \\
 $60-65$ & $29.8  \pm 4.1$ & $1.05 \times 10^{-3} \pm 0.29 \times 10^{-3}$ & $6.64 \times 10^{-2} \pm 0.19 \times 10^{-2}$ & $0.08$ \\
\end{tabular}
\end{ruledtabular}
\label{tab:chi}
\end{table*}

It should be emphasized that the parametrization in Eq.~(\ref{eq:nomc2}) is
practically necessary. The $\beta$ parameter can absorb any effects independent of
pseudorapidity space correlations. For a wider centrality bin, 
the width of the multiplicity distribution becomes broader,
since events with a wider range of centralities are included in the bin.
This causes the systematic difference of $\beta$ in the 5\% and
10\% centrality data sets as shown in Fig.~\ref{fig:bechi}~a).
The systematic shift of $\beta$ parameters to smaller values in the smaller centrality bin width,
suggests that $\beta$ dominantly contains fluctuations on \npart.
The ambiguity of \npart measured by PHENIX is not large compared, for 
example, to NA49 where a non-monotonic behavior of the scaled variance of 
multiplicities was seen
as a function of the number of projectile participant nucleons~\cite{NA49}.
In NA49, only spectators from the projectile nucleus are measurable,
causing an increase of scaled variance of multiplicity distributions in
peripheral collisions due to dominantly large \npart fluctuations 
in the target nucleus~\cite{NA49INTERPRET}.
This is due to the partial sampling with respect to the total number of nucleons
in two colliding nuclei. Since both projectile and target nuclei on both sides
can be measured by BBC and ZDC at PHENIX, such ambiguities of \npart are  
suppressed, even in peripheral collisions. 
Some \npart fluctuations remain, but the $\beta$ parameter can absorb 
this kind of fluctuation offset. Consequently, \npart
fluctuations are not harmful for the measurement of the $\alpha\xi$ products,
since they are based on the differential values of fluctuations for a given centrality bin.
In addition, $\beta$ is expected to absorb effects from azimuthal correlations.
Since the PHENIX detector does not cover the full azimuthal range, fluctuations
of low $p_T$ particles caused by reaction plane rotations and elliptic flow
should contribute to the two particle correlation function even in the
pseudorapidity direction as an offset in principle.
Owing to the $\beta$ parameter, the non-monotonic behavior of the
measured $\alpha\xi$ in the pseudorapidity direction
cannot be biased by elliptic flow nor by initial geometrical biases,
since the azimuthal correlations are constant over the narrow pseudorapidity window of
$|\eta|$~$<$~0.35~\cite{V2}.

\section{DISCUSSION} 
\label{sec:discussion}
\subsection{Other correlation sources}
\label{subsec:other}
We discuss three other sources of correlation
which are not related to density correlations we are interested in,
but could affect the measurement of the inclusive 
charged particle multiplicity fluctuations.  The first is charged
track pairs from particle decays in flight. The second is background charged
track pairs originating from secondary particle interactions
in  detector materials (i.e. showers, conversion pairs).
For these two sources we have estimated the effects of contaminations to 
the inclusive charged particle multiplicity fluctuations 
by GEANT-based MC~\cite{GEANT} simulations.
The third source is the known short-range correlation 
due to Bose-Einstein correlation of identical particles. 

The detectable charged particle compositions in the no magnetic field condition with
the selection criteria of charged tracks in Sec.~\ref{subsec:track} are estimated as
94\% for charged pions, 4\% for charged kaons and 2\% for proton and antiproton
in 0~-~70\% centrality. These are obtained by MC simulations based on identified
charged particle spectra measured by the PHENIX~\cite{PPG026} up to 4~GeV/$c$ of transverse
momentum, $p_T$. The statistically dominant weak decay particles which can contribute to the
inclusive charged particle multiplicity are $K^{0}_S$~$\to$~$\pi^{+}\pi^{-}$ and
$\Lambda$(\lbar)~$\to$~$p(\pbar)\pi^{-}(\pi^{+})$.
The relative invariant yields of those particles to charged pions
are 15\% and 5\% for $K^{0}_S$ and $\Lambda(\lbar)$~\cite{LAMBDA}, respectively.
They were calculated by the measured production cross section in Au+Au collisions
at \snn~$=$~200~GeV. The production cross section of $K^{0}_S$ is assumed to
be same as charged kaons~\cite{PPG026}. The detection efficiency of the charged track pairs
from weak decay particles in the one arm acceptance of PHENIX detector
($|\eta|$~$<$~0.35, $\Delta\phi$~$<$~$\pi/2$) is obtained by the MC simulation.
We estimated it by using the $p_T$ spectra of charged kaons for $K^{0}_S$ as the most
dominant meson, and by using the $p_T$ spectra of charged pions with transverse mass scaling
for $\Lambda$(\lbar) as the most dominant baryon, which contribute to the inclusive charged
particle multiplicity fluctuation.
As the result, the ratios of charged track pairs originating from those weak decay particles
to the number of produced charged pions per event are 0.7\% and 0.9\%
for $K^{0}_S$ and $\Lambda+\lbar$, respectively.
The effects of those correlations on $k$ were estimated as follows.
Suppose two independent NBD's in different windows have the same NBD
parameters of $\mu$ and $k$ for a given window size of $\delta\eta/2$.
If there is no correlation between the two windows, NBD in the $\delta\eta$
window size becomes a convoluted distribution between the two NBD's.
This is certainly true, since we know the correlation length is well below
the minimum size of $\delta\eta$ windows as already discussed.
Based on the NBD convolution theorem,
the convoluted NBD parameters, $\mu_{conv}$ and $k_{conv}$ are expressed as
$\mu_{conv}=2\mu$ and $k_{conv}=2k$ respectively in the case of no correlation.
For the case where the correlated pairs are embedded,
we define the fraction of the number of correlated pairs with respected to $\mu$ as $f$.
Then the mean value before the correlated pairs are embedded
is expressed as $\mu(1-f)$ in the $\delta\eta/2$ window.
The effect of the embedded correlation on $k_{conv}$ can be estimated 
by adding the number of correlated pairs to both windows simultaneously 
with the fraction of $f$. With $\mu(1-f)$ and $k$, we can generate NBD
with a random number generator in each window of $\delta\eta/2$ and convolute the two NBD's.
From the NBD fit to the convoluted distribution, we can obtain $k_{conv}$ including
the effect of the correlated pairs. We define the ratio of the deviation of $k_{conv}$
to the independent case, $\Delta k \equiv (k_{conv}-2k)/2k$
for $K^{0}_S$ and $\Lambda+\lbar$, respectively.
For all observed $(\langle\mu_c\rangle, \langle k_c\rangle)$ values
in all $\delta\eta$ windows in all centralities,
we have estimated $\Delta k$. The pair fraction, $f$ depends on $\delta\eta$ window size,
since weak decay particles have their own correlation length due to the kinematical constraint. 
The fraction $f$'s were obtained based on the two particle correlation
of decayed pairs as a function of $\delta\eta$ window size
which were evaluated from the GEANT-based MC simulation
with the actual track selection criteria. It should be noted that the integrated fractions
correspond to the above mentioned fractions, 0.7\% and 0.9\%
for $K^{0}_S$ and $\Lambda+\lbar$, respectively.
As the result, the average values of $\Delta k$ over all data points were estimated as
$+$~0.27\%~$\pm$~0.35\%~(standard deviation) and $+$~0.40\%~$\pm$~0.35\%~(standard deviation)
for $K^{0}_S$ and $\Lambda+\lbar$ decays, respectively.
On the other hand, the average value of relative errors, 
$\delta\langle k_c\rangle (total) / \langle k_c\rangle$
in measured $k$ is $\pm$~7.34\%~$\pm$~3.29\%~(standard deviation). We confirmed that
the estimated $\Delta k$ values are all included within the range of the relative errors on measured $k$.
Therefore, we can conclude that the effect of the statistically dominant weak decay pairs 
with a few percent level on the $\alpha\xi$ product can not exceed the full error sizes 
of the $\alpha\xi$ products in Table~\ref{tab:chi}.
 
The amount of material before the tracking system is 1.3\% of a radiation length.
It produces electron-positron pairs with 1.0\% photon conversion probability.
Almost 100\% of photons up to 4~GeV/$c$ of $p_T$ are produced by decays from 
neutral pions.  The detection efficiency of electron-positron pairs which survive
after the requirement of the charged track associations and two track separations 
in Sec.~\ref{subsec:track} is estimated as 0.22\%. 
It was estimated by the MC simulations with flat $p_T$ distribution of photons.  
Since the opening angle of the conversion pairs are very small, 
these conversion electrons are strongly suppressed by the two track separation cuts.
Consequently, electron-positron pairs of $2.2$~$\times$~$10^{-3}$\% 
with respect to the produced charged pions per event, contribute to the multiplicity fluctuations.
The efficiency of charged track pairs, which is produced by the materials from single charged
hadrons as knock-on electrons (positrons), is estimated as less than $5.8$~$\times$~$10^{-5}$\%.
Since the total pair fractions are much smaller than that in weak decays by several orders
of magnitude, we can conclude that the effect of those secondary particles on the $\alpha\xi$
products are negligible.

If the observed correlation were to originate only from the Bose-Einstein effect,
then we would expect $\alpha$ to be directly related to the chaoticity parameter,
$\lambda$ in HBT analysis which is measured in relative momentum space, $q$.
A similar measurement in pseudorapidity space based on Eq.~(\ref{eq:nomc2})
in low energy $A+A$ collisions~\cite{BECORR},
indicates the direct relation between $\lambda$ and $\alpha$.
The observed two particle correlation strength $\alpha$ in pseudorapidity space
is weaker than $\lambda$ measured in $q$ space and essentially becomes zero 
for the particle pairs selected in the higher $q$ region where HBT effect also 
becomes zero.
This indicates that the observed pseudorapidity correlations in the 
lower energy $A+A$ collisions are essentially explained purely by the HBT effect.
In Au+Au collisions at \snn~$=$~$200$~GeV, measured $\lambda$ shows 
constant behavior as a function of \npart~within 12\% and
a monotonic \npart~ dependence of HBT radii has been
observed~\cite{ENOKIZONO, HBT}.
This implies that the non-monotonic behavior of the $\alpha\xi$
product can not be explained solely as a result of the known HBT effect, 
because $\alpha \propto \lambda$ is expected to be constant for any 
\npart~and $\xi$ which would be related to the HBT source radii
is expected to be monotonic, if the known HBT effect is
the only source of the correlation.
 
\subsection{Evaluation of the non-monotonic behavior of $\alpha\xi$}
\label{subsec:sig}
\begin{figure*}[tbh]
\includegraphics[width=1.0\linewidth]{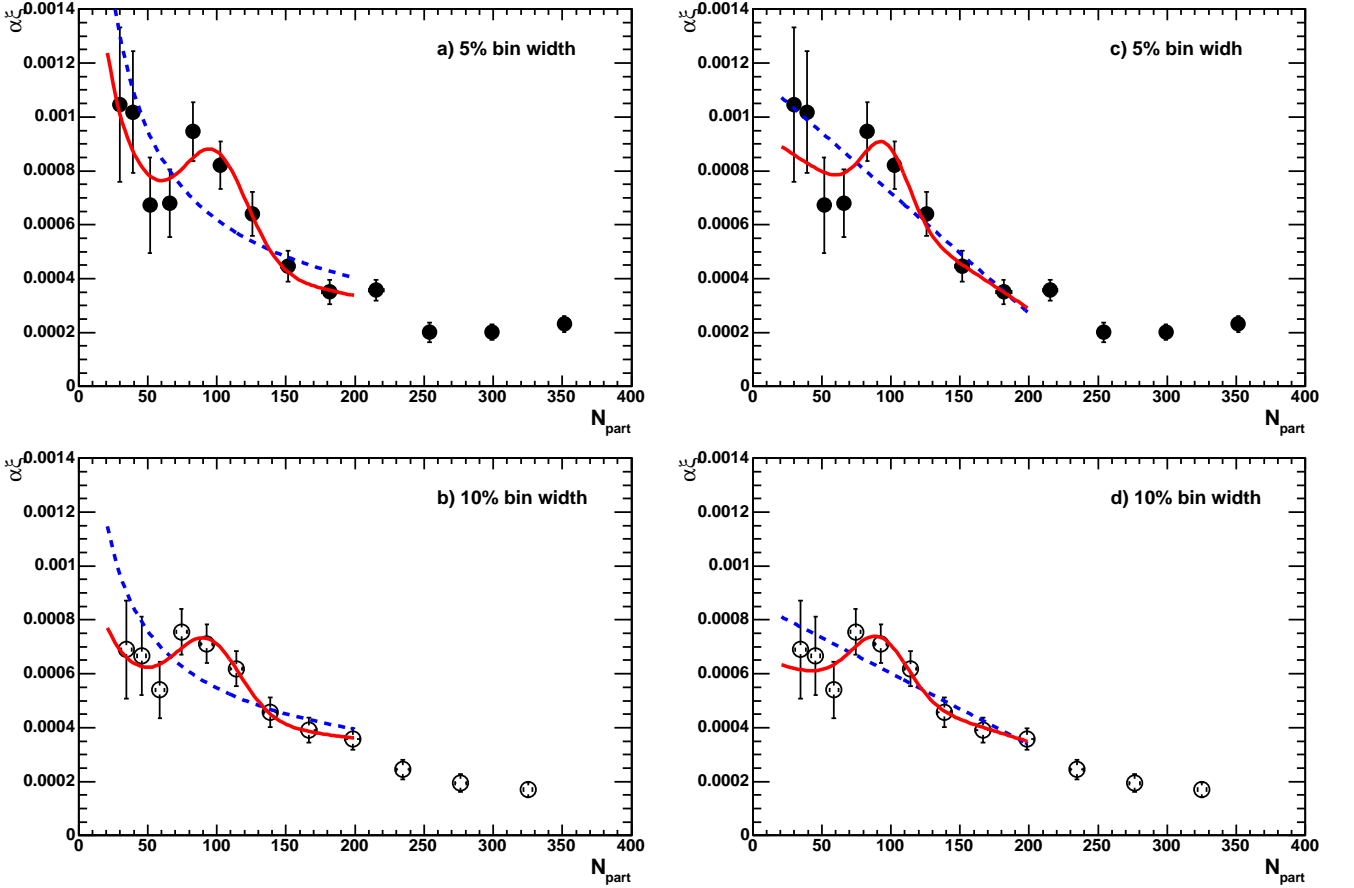}
\caption{
(Color online)
$\alpha\xi$ versus \npart~in Table~\ref{tab:chi} with fit curves.
The dashed and solid curves show the fit results with the baseline
functions Eq. 14 and 15 and with the composite functions Eq. 16
and 17 respectively.
a) and b) correspond to 5\% and 10\% bin width cases with the
power law baselines.
c) and d) correspond to 5\% and 10\% bin width cases with the
linear baselines.
}
\label{fig:mono}
\end{figure*}
\begin{table*}[tbh]
\caption{
The fit parameters in Eq.~(\ref{eq:pow}), Eq.~(\ref{eq:lin}), Eq.~(\ref{eq:comppow}) and Eq.~(\ref{eq:complin}).
}
\footnotesize
\begin{ruledtabular}
\begin{tabular}{lcccc}
Functional form & Centrality bin width (\%) & $\chi^2 / NDF (NDF)$ & $a \pm \delta a$ & Significance $(a / \delta a)$ \\
\hline
Power law in Eq.~(\ref{eq:pow})                & $5$  & $2.76 (7)$ &                                         &         \\ 
Power law + Gaussian in Eq.~(\ref{eq:comppow}) & $5$  & $0.60 (4)$ & $0.37 \times 10^3 \pm 0.09 \times 10^3$ & $3.98$  \\
Linear in Eq.~(\ref{eq:lin})                   & $5$  & $1.23 (7)$ &                                         &         \\ 
Linear + Gaussian in Eq.~(\ref{eq:complin})    & $5$  & $0.79 (4)$ & $0.27 \times 10^3 \pm 0.21 \times 10^3$ & $1.24$  \\
\hline
Power law in Eq.~(\ref{eq:pow})                & $10$ & $2.10 (7)$ &                                         &         \\ 
Power law + Gaussian in Eq.~(\ref{eq:comppow}) & $10$ & $0.38 (4)$ & $0.27 \times 10^3 \pm 0.08 \times 10^3$ & $3.21$  \\
Linear in Eq.~(\ref{eq:lin})                   & $10$ & $1.09 (7)$ &                                         &         \\ 
Linear + Gaussian in Eq.~(\ref{eq:complin})    & $10$ & $0.43 (4)$ & $0.22 \times 10^3 \pm 0.13 \times 10^3$ & $1.69$  \\
\end{tabular}
\end{ruledtabular}
\label{tab:test}
\end{table*}
\begin{figure}[tbh]
\includegraphics[width=1.0\linewidth]{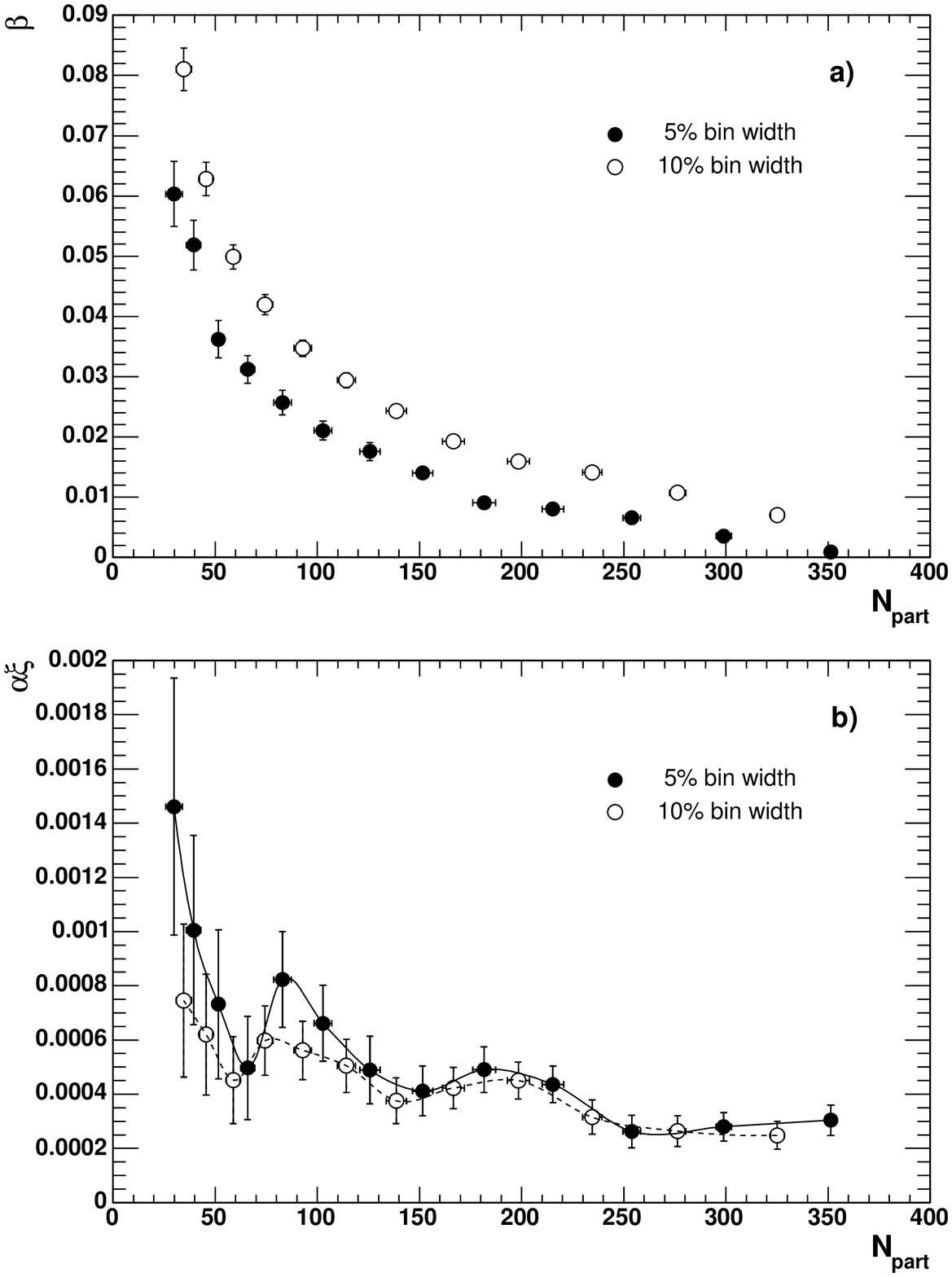}
\caption{
Fit results based on Eq.~(\ref{eq:app}) by limiting the range of $\delta\eta$~from
0.066 to 0.306. a) is $\beta$ and b) is products of $\alpha\xi$ as a function of \npart.
The horizontal error bars correspond to ambiguities in the mean value of~\npart~as
explained in Sec.~\ref{subsec:cent}.
The vertical error bars are obtained from errors on the fitting parameter.
}
\label{fig:parrange}
\end{figure}
The $\alpha\xi$ product obtained by Eq.~(\ref{eq:app}) is related
to susceptibility in the long wavelength limit, $\chi_{\omega=0}$
as described in Sec.~\ref{sec:observable}.
According to Eq.~(\ref{eq:alphaxi}), if the system temperature $T$ 
is far from the critical temperature $T_{c}$ then $\alpha\xi$ is expected to
decrease monotonically with increasing $T$ which is a monotonic function
of \npart~as will be discussed in Sec.~\ref{subsec:temp}.
Therefore, one can assume a monotonically decreasing function as a
hypothesis of the baseline in $T$ far from $T_{c}$.
As baseline functional forms for $\alpha\xi$ versus $T$ 
we consider the following two cases.
The first is a power law function which is naturally expected from 
Eq.~(\ref{eq:alphaxi}), and the second is a linear function as the
simplest assumption.
The power law baseline and the linear baseline are parametrized as
\begin{eqnarray}
\label{eq:pow}
\alpha \xi (\npart) = p_1 (\npart)^{p_2}
\end{eqnarray}
and 
\begin{eqnarray}
\label{eq:lin}
\alpha \xi (\npart) = p_1 + p_2 \npart
\end{eqnarray}
with fit parameter $p_1$ and $p_2$, respectively.
As a test hypothesis, we assume a local maximum on the monotonic
baselines in $\alpha\xi$ versus \npart.
Although the functional form around the local maximum is not known 
a priori without introducing a physical model, we can at least discuss
the significance of the local maximum above the monotonic baseline by
introducing a Gaussian distribution.
The composite functions are defined as 
\begin{eqnarray}
\label{eq:comppow}
\alpha \xi (N_{part}) = p_1 (N_{part})^{p_2} + a e^{-\frac{(N_{part} - m)^{2}}{2 w^{2}}}
\end{eqnarray}
and 
\begin{eqnarray}
\label{eq:complin}
\alpha \xi (N_{part}) = p_1 + p_2 \npart + a e^{-\frac{(N_{part} - m)^{2}}{2 w^{2}}},
\end{eqnarray}
where $a$, $m$ and $w$ correspond to amplitude, mean and width of
the Gaussian component, respectively.
Fits with the four functional forms were performed to $\alpha\xi$
versus \npart~in 20~$<$~\npart~$<$~200.
Figure~\ref{fig:mono} shows $\alpha\xi$ versus \npart~from Table~\ref{tab:chi}
with those fit curves.
The dashed and solid curves show the fit results with the baseline functions
and with the composite functions.
Figures~\ref{fig:mono}~a) and b) correspond to 5\% and 10\% bin width cases with the
power law baselines.
Figures~\ref{fig:mono}~c) and d) correspond to 5\% and 10\% bin width cases with the
linear baselines.
Table~\ref{tab:test} summarizes all the fit results from
the Minuit program~\cite{MINUIT}, including
functional forms, centrality bin widths, $\chi^2/NDF~(NDF)$, the Gaussian
amplitude $a$ with its error $\delta a$, and the significance of the amplitude
defined as $a / \delta a$.
Although the significance of the local maximum with the linear 
baseline is smaller than that with the power law baseline, 
this is mainly due to the larger uncertainty on $a$
in Eq.~(\ref{eq:complin}) than in Eq.~(\ref{eq:comppow}).
This reflects the fact that the combination of a Gaussian distribution 
with the linear baseline is not as good a fit
as that with the power law baseline for the given data points.
The difference on the significance between 5\% and 10\% centrality bin
width cases is not attributed to the correlations with $\beta$ parameter,
since $\beta$ was introduced as a parameter independent of $\delta\eta$.
This can, rather, be understood as the smearing effect of a 
peak-like shape due to the
difference of centrality bin widths around the mean~\npart.
In all cases in Table~\ref{tab:test}, the $\chi^2/NDF$ indicate that 
composite functions are favored over monotonic functions.
This result supports the non-monotonicity of $\alpha\xi$ as a function of \npart.

Although there is a possibility that more than one correlation length scale is
dynamically present, the functional form with one correlation length can reasonably
describe the region of 0.066~$<$~$\delta\eta$~$<$~0.7 with the average $\chi^2/NDF$
of 0.43 over all centralities.
We have performed a further check on the \npart~dependence of the
$\alpha\xi$ products by limiting the region of fit to 0.066~$<$~$\delta\eta$~$<$~0.306
as shown in Fig.~\ref{fig:parrange}.
The characteristic behavior of $\alpha\xi$ still present at around \npart~$\sim$~90.

\subsection{Initial temperature and \npart}
\label{subsec:temp}
The Bjorken energy density~\cite{BJ} derived from the measured $dE_T/d\eta$
in the mid-rapidity region is defined as
\begin{eqnarray}
\label{eq:njed}
\epsilon_{Bj} = \frac{1}{c \tau A_T} \frac{dE_T}{dy},
\end{eqnarray}
where $\tau$ is the formation time and $A_T$ is the nuclei transverse overlap
area size. It is well known that the energy density monotonically increases
with increasing \npart~in Au+Au collisions at \snn~$=$200~GeV~\cite{PPG019}.

This indicates that the change of \npart~even at the fixed collision energy
can yield a fine scan over the initial temperature.
Therefore, the non-monotonic behavior in the $\alpha\xi$ products
could be an indication of the critical initial temperature as defined in
Eq.(\ref{eq:sus}).
The Bjorken energy density, $\epsilon_{Bj}\tau$, at the local maximum of
$\alpha\xi$ seen at \npart~$\sim$~90 corresponds to
$\sim$~2.4~GeV/(fm$^{2}\cdot c)$ with $A_T=60$~fm$^{2}$.

\section{CONCLUSIONS}
\label{sec:conclusion}
The multiplicity distributions measured in Au+Au collisions at \snn~$=$~200~GeV
are found to be well-described by negative binomial distributions.

The two-particle density correlations have been measured via the functional
form for  pseudorapidity density fluctuations derived in the Ginzburg-Landau
framework, up to the second order term in the free energy, with a scalar order
parameter defined as pseudorapidity-dependent multiplicity fluctuations around
the mean value.
The functional form can reasonably fit $k$~versus~$\delta\eta$ in all centralities
in $|\eta|$~$<$~0.35 region with one correlation length assumption and the constant
term $\beta$ which is independent of $\delta\eta$.
We found $\beta$ is necessary to absorb residual effects of finite centrality binning.

We found that the absolute scale of the correlation length, $\xi$ 
depends on the magnitude of the correlation strength at zero distance, $\alpha$
within the range of pseudorapidity window sizes available in this analysis.
However, according to the free parameter fit results, the upper limit on
$\xi$~$<$~0.035 was obtained, and it was confirmed by the accuracy of the fits
with approximated integrated correlation function in the limit of
the small correlation length ($\xi \ll \delta\eta$).

The $\alpha\xi$ product in the correlation function, which is monotonically 
related to susceptibility in the long wavelength limit $\chi_{\omega=0}$, was
seen to exhibit a non-monotonic behavior
as a function of the number of participant nucleons~\npart.
A possible indication of a local maximum is seen at \npart~$\sim$~90 and the
corresponding energy density based on the Bjorken picture is
$\epsilon_{Bj}\tau$~$\sim$~2.4~GeV/(fm$^{2}\cdot c)$ with the transverse area
size of 60~fm$^{2}$.

Trivial particle correlations originating from charged track reconstructions
in tracking detectors have been suppressed in this analysis a priori.
The ratio of charged particles from statistically dominant weak decays
and secondary particles produced in the detector materials, which
contribute as correlated pairs, are respectively estimated below
$\sim 1$\% and $\sim 10^{-3}$\% with respect to the total number of
charged pions in the PHENIX acceptance per event.
We have estimated those effects on measured $k$ values and the deviation due to
the effect is well inside the total error size of observed $\delta \langle k_c\rangle$.
Therefore, we conclude that their contributions are almost negligible to 
the observed behavior of the $\alpha\xi$ products.

The behavior may be explained by the onset of a mixture of different types of particle
production mechanisms which are not necessarily related to temperature 
or density
correlations.  However, interpreted within the Ginzburg-Landau framework
the local maximum of the $\alpha\xi$ product could be an indication
of a critical phase boundary.

\begin{acknowledgments}
((Temporally))
We thank the staff of the Collider-Accelerator and Physics
Departments at Brookhaven National Laboratory and the staff
of the other PHENIX participating institutions for their
vital contributions.  We acknowledge support from the
Department of Energy, Office of Science, Nuclear Physics
Division, the National Science Foundation, Abilene Christian
University Research Council, Research Foundation of SUNY,
and Dean of the College of Arts and Sciences, Vanderbilt
University (U.S.A), Ministry of Education, Culture, Sports,
Science, and Technology and the Japan Society for the
Promotion of Science (Japan), Conselho Nacional de
Desenvolvimento Cient\'{\i}fico e Tecnol{\'o}gico and
Funda\c c{\~a}o de Amparo {\`a} Pesquisa do Estado de
S{\~a}o Paulo (Brazil), Natural Science Foundation of China
(People's Republic of China), Centre National de la
Recherche Scientifique, Commissariat {\`a} l'{\'E}nergie
Atomique, Institut National de Physique Nucl{\'e}aire et
de Physique des Particules, and Institut National
de Physique Nucl{\'e}aire et de Physique des Particules,
(France), Bundesministerium fuer Bildung und
Forschung, Deutscher Akademischer Austausch Dienst, and
Alexander von Humboldt Stiftung (Germany), Hungarian
National Science Fund, OTKA (Hungary), Department of Atomic
Energy and Department of Science and Technology (India),
Israel Science Foundation (Israel), Korea Research
Foundation and Center for High Energy Physics (Korea),
Russian Ministry of Industry, Science and Tekhnologies,
Russian Academy of Science, Russian Ministry of Atomic Energy
(Russia), VR and the Wallenberg Foundation (Sweden), the U.S.
Civilian Research and Development Foundation for the Independent
States of the Former Soviet Union, the US-Hungarian 
NSF-OTKA-MTA,
the US-Israel Binational Science Foundation, and the 5th
European Union TMR Marie-Curie Programme.
\end{acknowledgments}

\begin{appendix}
\section{Definition of correlation length and susceptibility in Ginzburg-Landau framework}
\label{ap:deri}
The Ginzburg-Landau~(GL)~\cite{GL} framework with the Ornstein-Zernike
picture~\cite{OZ} for a scalar order parameter is briefly reviewed.
The relations
with correlation length and susceptibility are explicitly derived in this appendix.

The first attempt to apply free energy discussions to nucleus-nucleus collisions
can be found in~\cite{FREEENERGY}; application to the QCD phase
transition is in~\cite{DENSITY}.
GL describes the relation between a free energy density $f$ and an order parameter
$\phi$ as a function of the system temperature~$T$.
By adding a spatially inhomogeneous term~$A(T)(\nabla\phi)^2$ and an external
field $h$, the general form is described as follows
\begin{eqnarray}
\label{eq:free}
f(T,\phi,h)= f_0(T)+ \frac{1}{2}A(T)(\nabla\phi)^2+ \nonumber \\
\frac{1}{2}a(T)\phi^2 +\frac{1}{4}b\phi^4+ \cdot\cdot\cdot -h\phi,
\end{eqnarray}
where $f_0$ is the equilibrium value of the free energy, terms with odd powers
are neglected due to the symmetry of the order parameter, and the sign of $b$ is
used to classify the transition orders: $b$~$<$~0 for the first order, $b$~$>$~0
for the second order and $b$~$=$~0 at the critical point.
Since the order parameter should vanish above a critical temperature $T_c$,
it is natural for the coefficient $a(T)$ to be expressed as $a(T)$~$=$~$a_0|T-T_c|$,
while $b$ is usually assumed to be constant in the vicinity of $T_c$.
In the following, we neglect higher order terms beyond the second order term.
This approximation corresponds to a picture where a system approaches 
the phase boundary 
from afar, since $\phi$ is close to zero in the regions far from $T_c$. 
In this sense, the approximation is insensitive to the details of the phase
transition order, {\it i.e.} higher order terms, 
but only sensitive to the behavior near $T_c$.

We apply this GL framework to density correlations in 
the longitudinal space coordinate $z$ in heavy-ion collisions.
The system of the produced matter dynamically evolves, so
we introduce the proper time frame for each sub element.
The longitudinal space element becomes $dz$~=~$\tau$~cosh$(y)dy$,
at a fixed proper time $\tau$, where $y$ is rapidity, as introduced in~\cite{FRAME}.
Since we measure the density fluctuations in the mid-rapidity regions
$|\eta|$~$<$~0.35 as described in Sec.~\ref{sec:detector}, we use $dy$ in
place of $dz$ by the approximation of cosh$(y)$~=~1 to simplify the form
of the correlation function derived from GL free energy.

The order parameter of this analysis corresponds to multiplicity 
density fluctuations of inclusive charged particles around the mean density.
Fluctuations are measured as a function of a one-dimensional 
pseudorapidity point~$\eta$, defined as
\begin{eqnarray}
\label{eq:order}
\phi(\eta) = \rho(\eta) - \langle\rho(\eta)\rangle,
\end{eqnarray}
where the pair of brackets indicates an operator to take the average.
In the above mentioned rapidity region, rapidity can be represented
by pseudorapidity to a good approximation, as explained in the last
paragraph of Sec.~\ref{subsec:track}.

With the Fourier expansion of the density fluctuation at 
pseudorapidity point~$\eta$, 
$\phi(\eta) = \sum_{\omega}{\phi_{\omega} e^{i\omega \eta}}$, where $\omega$
is wave number, one can express the deviation of the free energy density
due to spatial fluctuations from the equilibrium value $f_0$
\begin{eqnarray}
\label{eq:devfree}
\Delta F/Y &=& \frac{1}{Y}\int (f-f_0) d\eta \nonumber\\
           &=& \frac{1}{2}\sum_{\omega}{|\phi_{\omega}|^2(a(T)+A(T)\omega^{2})},
\end{eqnarray}
where $Y$ is the total pseudorapidity range corresponding to a 
one-dimensional volume. Terms
up to the second order are included in the approximation
in the vicinity of the critical point in Eq.~(\ref{eq:free}).
Given the free energy deviation, one can obtain the statistical
weight~$W$ for fluctuation~$\phi(\eta)$ to occur in a given temperature~$T$
\begin{eqnarray}
\label{eq:prob}
W(\phi(\eta)) = N e^{-\Delta F/ T}.
\end{eqnarray}
Therefore the statistical average of the square of the density fluctuation
with the wave number~$\omega$ is described as
\begin{eqnarray}
\label{eq:weight}
\langle |\phi_{\omega}|^2 \rangle &=&
\int_{-\infty}^{+\infty} |\phi_{\omega}|^2 W\left(\sum_{\omega}{\phi_{\omega} e^{i \omega \eta}}\right) d\phi_{\omega} \nonumber \\
                           &=& \frac{NT}{Y}\frac{1}{a(T)+A(T)\omega^2}.
\end{eqnarray}

An experimentally observable two point density correlation function can be
related to the statistical average of the square of the density fluctuation.
With a density~$\rho(\eta_{i})$ for a given sub-volume~$d\eta_{i}$,
the two point density correlation~$G_2$ is expressed in the case of
$\langle \rho(\eta_{1}) \rangle = \langle \rho(\eta_{2}) \rangle = \langle \rho \rangle $ as
\begin{eqnarray}
\label{eq:g2}
G_2(\eta_{1}, \eta_{2}) = \langle (\rho(\eta_{1})- \langle \rho \rangle)(\rho(\eta_{2})- \langle \rho \rangle) \rangle,
\end{eqnarray}
where case $1$ coinciding with case $2$ is excluded to
simplify the following discussion.
Multiplying both sides of Eq.~(\ref{eq:g2}) by $e^{-i \omega \eta}$~$\equiv$~$e^{-i \omega (\eta_{2}-\eta_{1})}$
and integrating over sub-volume~$d\eta_{1}$ and $d\eta_{2}$ gives
\begin{eqnarray}
\label{eq:intg2}
Y\int G_2(\eta) e^{-i \omega \eta}d\eta &=&
\langle | \int (\rho(\eta)- \langle \rho \rangle ) e^{-i \omega \eta} d\eta |^2 \rangle \nonumber \\
&=& \langle |\phi_\omega|^2 \rangle.
\end{eqnarray}
From Eq.~(\ref{eq:weight}) and (\ref{eq:intg2}),
$G_2$ can be related to the inverse Fourier transformation
of the statistical average of~$|\phi_\omega|^2$.
Therefore in the one-dimensional case $G_2$ is described as
\begin{eqnarray}
\label{eq:fg2}
G_2(\eta) = \frac{NT}{2Y^2 A(T)} \xi(T) e^{-|\eta|/\xi(T)},
\end{eqnarray}
where the correlation length~$\xi(T)$ is introduced, which is defined as
\begin{eqnarray}
\label{eq:corrlength}
\xi(T)^2 = \frac{A(T)}{a_0|T-T_c|}.
\end{eqnarray}
In general, a singular behavior of $\xi(T)$ as a function of $T$
indicates the critical temperature of the phase transition.

The wave number dependent susceptibility can also be defined from 
the expansion of the GL free energy based on Eq.~(\ref{eq:free})
and (\ref{eq:devfree}) as follows,
\begin{eqnarray}
\label{eq:chi}
\chi_{\omega} &=& -\left(\frac{\partial^{2}f}{\partial h^{2}}\right)
                  = \left(\frac{\partial h}{\partial \phi_{\omega}}\right)^{-1} \nonumber \\
              &=& \left(\frac{\partial^{2}(\Delta F/Y)}{\partial \phi_{\omega}^{2}}\right)^{-1}  \nonumber \\
              &=& \frac{1}{a_{0}|T-T_{c}|(1 + \omega^{2} \xi (T)^2)}.
\end{eqnarray}
In the case of the long wavelength limit of $\omega$~$=$~0, the susceptibility can be expressed as, 
\begin{eqnarray}
\label{eq:sus}
\chi_{\omega = 0} = \frac{1}{a_{0}|T-T_{c}|} = \frac{2Y^{2}}{NT}\xi(T)G_{2}(0).
\end{eqnarray}
In this framework, the $\xi$ and $\chi_{\omega=0}$ diverge at the same temperature.

\section{Tables of NBD fit results}
\label{ap:table}
NBD fit results in all window sizes in all centrality bins in Fig.~\ref{fig:kdetachi}.
In the following
Table~\ref{ap_01} through Table~\ref{ap_12} and Table~\ref{ap_13} through Table~\ref{ap_25}
correspond to results in 10\% and 5\% bin width cases, respectively.
$\langle \mu_c\rangle$ and $\langle \mu\rangle$ are weighted means of corrected and uncorrected
$\mu$ over all window positions respectively.
$\langle k_c\rangle$ and $\langle k\rangle$ are weighted means of corrected and uncorrected
$k$ over all window positions respectively.
Statistical errors on weighted means $\delta\langle k_c\rangle$~(stat)
are obtained as explained in Sec.~\ref{subsec:errors}.
$\langle \chi^2/NDF\rangle$ is the average of reduced $\chi^2$ of NBD fits over all window positions.
$\langle NDF\rangle$ is the average of the degree of freedom of NBD fits over all window positions.
Systematic errors $\delta\langle k_c\rangle$~(dead), $\delta\langle k_c\rangle~$(fake) and
$\delta\langle k_c\rangle$~(total) are explained in Sec.~\ref{subsec:errors}.

\begingroup \squeezetable

\begin{table*}[tbh]
\caption{NBD fit results in centrality 0-10\%.
}
\begin{ruledtabular}
\begin{tabular}{ccccccc}
 $\delta\eta$ & $\langle \mu_{c} \rangle (\langle \mu \rangle)$ & $\langle k_{c} \rangle (\langle k \rangle)$ & $\langle \chi^{2}/NDF \rangle (\langle NDF \rangle)$ & $\delta \langle k_{c} \rangle (dead)$ & $\delta \langle k_{c} \rangle (fake)$ & $\delta \langle k_{c} \rangle (total)$ \\
\hline
 $0.700$ & $77.535 (60.011) \pm 0.108$ & $114.28 (88.45) \pm 3.21$ & $0.96 (71.0)$ & $\pm 2.89$ & $\pm 2.28$ & $\pm 4.89$\\
 $0.678$ & $75.445 (58.445) \pm 0.106$ & $113.11 (87.62) \pm 3.29$ & $0.89 (69.7)$ & $\pm 4.92$ & $\pm 3.31$ & $\pm 6.78$\\
 $0.656$ & $72.977 (56.455) \pm 0.103$ & $114.40 (88.50) \pm 3.38$ & $0.95 (68.2)$ & $\pm 4.33$ & $\pm 3.04$ & $\pm 6.28$\\
 $0.634$ & $70.485 (54.450) \pm 0.101$ & $114.02 (88.08) \pm 3.42$ & $0.92 (66.7)$ & $\pm 4.54$ & $\pm 3.01$ & $\pm 6.43$\\
 $0.613$ & $67.998 (52.489) \pm 0.099$ & $114.44 (88.33) \pm 3.52$ & $0.94 (65.3)$ & $\pm 3.79$ & $\pm 2.31$ & $\pm 5.66$\\
 $0.591$ & $65.530 (50.572) \pm 0.096$ & $114.28 (88.18) \pm 3.60$ & $0.95 (63.9)$ & $\pm 3.59$ & $\pm 2.62$ & $\pm 5.72$\\
 $0.569$ & $63.050 (48.609) \pm 0.094$ & $114.62 (88.35) \pm 3.71$ & $0.97 (62.3)$ & $\pm 3.88$ & $\pm 2.80$ & $\pm 6.05$\\
\end{tabular}
\end{ruledtabular}
\label{ap_01}
\end{table*}

\endgroup
\begingroup \squeezetable

\begin{table*}[tbh]
\caption{NBD fit results in centrality 5-15\%.
}
\begin{ruledtabular}
\begin{tabular}{ccccccc}
 $\delta\eta$ & $\langle \mu_{c} \rangle (\langle \mu \rangle)$ & $\langle k_{c} \rangle (\langle k \rangle)$ & $\langle \chi^{2}/NDF \rangle (\langle NDF \rangle)$ & $\delta \langle k_{c} \rangle (dead)$ & $\delta \langle k_{c} \rangle (fake)$ & $\delta \langle k_{c} \rangle (total)$ \\
\hline
 $0.700$ & $66.445 (51.428) \pm 0.098$ & $80.80 (62.54) \pm 1.92$ & $1.20 (68.0)$ & $\pm 2.14$ & $\pm 2.17$ & $\pm 3.60$\\
 $0.678$ & $64.658 (50.089) \pm 0.097$ & $80.40 (62.28) \pm 1.95$ & $1.15 (66.4)$ & $\pm 3.40$ & $\pm 1.53$ & $\pm 4.21$\\
 $0.656$ & $62.535 (48.377) \pm 0.094$ & $81.17 (62.79) \pm 1.99$ & $1.20 (65.0)$ & $\pm 2.28$ & $\pm 1.11$ & $\pm 3.22$\\
 $0.634$ & $60.379 (46.644) \pm 0.092$ & $81.10 (62.65) \pm 2.02$ & $1.23 (64.1)$ & $\pm 2.24$ & $\pm 1.57$ & $\pm 3.40$\\
 $0.613$ & $58.258 (44.970) \pm 0.090$ & $80.61 (62.21) \pm 2.06$ & $1.21 (62.4)$ & $\pm 2.75$ & $\pm 1.43$ & $\pm 3.72$\\
 $0.591$ & $56.156 (43.337) \pm 0.088$ & $79.66 (61.47) \pm 2.08$ & $1.12 (60.6)$ & $\pm 2.85$ & $\pm 1.50$ & $\pm 3.83$\\
 $0.569$ & $54.043 (41.665) \pm 0.085$ & $79.26 (61.09) \pm 2.12$ & $1.07 (59.2)$ & $\pm 2.59$ & $\pm 1.88$ & $\pm 3.83$\\
\end{tabular}
\end{ruledtabular}
\label{ap_02}
\end{table*}

\endgroup
\begingroup \squeezetable

\begin{table*}[tbh]
\caption{NBD fit results in centrality 10-20\%.}
\begin{ruledtabular}
\begin{tabular}{ccccccc}
 $\delta\eta$ & $\langle \mu_{c} \rangle (\langle \mu \rangle)$ & $\langle k_{c} \rangle (\langle k \rangle)$ & $\langle \chi^{2}/NDF \rangle (\langle NDF \rangle)$ & $\delta \langle k_{c} \rangle (dead)$ & $\delta \langle k_{c} \rangle (fake)$ & $\delta \langle k_{c} \rangle (total)$ \\
\hline
 $0.700$ & $55.290 (42.794) \pm 0.090$ & $63.58 (49.21) \pm 1.47$ & $1.14 (64.0)$ & $\pm 0.37$ & $\pm 0.57$ & $\pm 1.62$\\
 $0.678$ & $53.832 (41.701) \pm 0.089$ & $62.10 (48.11) \pm 1.46$ & $0.93 (63.1)$ & $\pm 2.48$ & $\pm 2.60$ & $\pm 3.88$\\
 $0.656$ & $52.092 (40.299) \pm 0.087$ & $61.78 (47.79) \pm 1.48$ & $0.88 (62.3)$ & $\pm 2.34$ & $\pm 1.54$ & $\pm 3.17$\\
 $0.634$ & $50.301 (38.858) \pm 0.085$ & $61.91 (47.82) \pm 1.51$ & $0.90 (60.3)$ & $\pm 2.06$ & $\pm 1.23$ & $\pm 2.83$\\
 $0.613$ & $48.533 (37.463) \pm 0.082$ & $61.80 (47.70) \pm 1.54$ & $0.88 (58.5)$ & $\pm 2.10$ & $\pm 1.09$ & $\pm 2.83$\\
 $0.591$ & $46.769 (36.093) \pm 0.080$ & $62.00 (47.85) \pm 1.58$ & $0.90 (57.9)$ & $\pm 1.74$ & $\pm 1.21$ & $\pm 2.64$\\
 $0.569$ & $45.006 (34.698) \pm 0.078$ & $61.89 (47.71) \pm 1.61$ & $0.87 (56.6)$ & $\pm 1.77$ & $\pm 1.06$ & $\pm 2.62$\\
\end{tabular}
\end{ruledtabular}
\label{ap_03}
\end{table*}

\endgroup
\begingroup \squeezetable

\begin{table*}[tbh]
\caption{NBD fit results in centrality 15-25\%.}
\begin{ruledtabular}
\begin{tabular}{ccccccc}
 $\delta\eta$ & $\langle \mu_{c} \rangle (\langle \mu \rangle)$ & $\langle k_{c} \rangle (\langle k \rangle)$ & $\langle \chi^{2}/NDF \rangle (\langle NDF \rangle)$ & $\delta \langle k_{c} \rangle (dead)$ & $\delta \langle k_{c} \rangle (fake)$ & $\delta \langle k_{c} \rangle (total)$ \\
\hline
 $0.700$ & $45.868 (35.502) \pm 0.082$ & $53.43 (41.35) \pm 1.25$ & $1.06 (57.0)$ & $\pm 1.83$ & $\pm 1.78$ & $\pm 2.85$\\
 $0.678$ & $44.657 (34.594) \pm 0.080$ & $53.48 (41.43) \pm 1.29$ & $0.95 (55.3)$ & $\pm 1.49$ & $\pm 0.53$ & $\pm 2.04$\\
 $0.656$ & $43.211 (33.428) \pm 0.078$ & $53.45 (41.35) \pm 1.31$ & $0.98 (54.2)$ & $\pm 1.53$ & $\pm 0.71$ & $\pm 2.14$\\
 $0.634$ & $41.737 (32.242) \pm 0.076$ & $53.70 (41.48) \pm 1.35$ & $0.96 (53.0)$ & $\pm 1.43$ & $\pm 0.58$ & $\pm 2.05$\\
 $0.613$ & $40.288 (31.098) \pm 0.075$ & $53.41 (41.23) \pm 1.38$ & $0.92 (51.2)$ & $\pm 1.94$ & $\pm 1.12$ & $\pm 2.63$\\
 $0.591$ & $38.829 (29.965) \pm 0.073$ & $53.66 (41.42) \pm 1.42$ & $0.96 (50.3)$ & $\pm 1.71$ & $\pm 0.70$ & $\pm 2.33$\\
 $0.569$ & $37.368 (28.808) \pm 0.071$ & $53.48 (41.23) \pm 1.45$ & $0.94 (49.0)$ & $\pm 1.76$ & $\pm 0.82$ & $\pm 2.42$\\
\end{tabular}
\end{ruledtabular}
\label{ap_04}
\end{table*}

\endgroup
\begingroup \squeezetable

\begin{table*}[tbh]
\caption{NBD fit results in centrality 20-30\%.
}
\begin{ruledtabular}
\begin{tabular}{ccccccc}
 $\delta\eta$ & $\langle \mu_{c} \rangle (\langle \mu \rangle)$ & $\langle k_{c} \rangle (\langle k \rangle)$ & $\langle \chi^{2}/NDF \rangle (\langle NDF \rangle)$ & $\delta \langle k_{c} \rangle (dead)$ & $\delta \langle k_{c} \rangle (fake)$ & $\delta \langle k_{c} \rangle (total)$ \\
\hline
 $0.700$ & $37.610 (29.110) \pm 0.073$ & $47.51 (36.77) \pm 1.17$ & $0.82 (53.0)$ & $\pm 1.28$ & $\pm 0.76$ & $\pm 1.89$\\
 $0.678$ & $36.606 (28.358) \pm 0.072$ & $47.66 (36.92) \pm 1.20$ & $0.86 (51.7)$ & $\pm 1.61$ & $\pm 0.68$ & $\pm 2.12$\\
 $0.656$ & $35.403 (27.388) \pm 0.070$ & $47.76 (36.94) \pm 1.22$ & $0.83 (51.4)$ & $\pm 1.10$ & $\pm 0.43$ & $\pm 1.70$\\
 $0.634$ & $34.199 (26.419) \pm 0.069$ & $47.45 (36.66) \pm 1.24$ & $0.83 (50.2)$ & $\pm 1.54$ & $\pm 0.65$ & $\pm 2.08$\\
 $0.613$ & $33.002 (25.474) \pm 0.067$ & $47.80 (36.90) \pm 1.27$ & $0.97 (49.3)$ & $\pm 0.99$ & $\pm 0.68$ & $\pm 1.75$\\
 $0.591$ & $31.812 (24.550) \pm 0.065$ & $47.37 (36.56) \pm 1.28$ & $0.96 (48.6)$ & $\pm 1.29$ & $\pm 0.62$ & $\pm 1.92$\\
 $0.569$ & $30.609 (23.598) \pm 0.063$ & $47.42 (36.56) \pm 1.31$ & $0.93 (47.7)$ & $\pm 1.12$ & $\pm 0.51$ & $\pm 1.80$\\
\end{tabular}
\end{ruledtabular}
\label{ap_05}
\end{table*}

\endgroup
\begingroup \squeezetable

\begin{table*}[tbh]
\caption{NBD fit results in centrality 25-35\%.
}
\begin{ruledtabular}
\begin{tabular}{ccccccc}
 $\delta\eta$ & $\langle \mu_{c} \rangle (\langle \mu \rangle)$ & $\langle k_{c} \rangle (\langle k \rangle)$ & $\langle \chi^{2}/NDF \rangle (\langle NDF \rangle)$ & $\delta \langle k_{c} \rangle (dead)$ & $\delta \langle k_{c} \rangle (fake)$ & $\delta \langle k_{c} \rangle (total)$ \\
\hline
 $0.700$ & $30.796 (23.836) \pm 0.065$ & $41.09 (31.80) \pm 1.05$ & $0.77 (48.0)$ & $\pm 1.99$ & $\pm 0.38$ & $\pm 2.28$\\
 $0.678$ & $29.940 (23.193) \pm 0.063$ & $41.40 (32.07) \pm 1.06$ & $0.83 (48.9)$ & $\pm 1.53$ & $\pm 0.13$ & $\pm 1.87$\\
 $0.656$ & $28.970 (22.412) \pm 0.062$ & $41.18 (31.85) \pm 1.07$ & $0.81 (47.9)$ & $\pm 1.52$ & $\pm 0.55$ & $\pm 1.94$\\
 $0.634$ & $27.967 (21.605) \pm 0.060$ & $41.07 (31.73) \pm 1.09$ & $0.84 (46.8)$ & $\pm 1.36$ & $\pm 0.47$ & $\pm 1.81$\\
 $0.613$ & $26.977 (20.823) \pm 0.059$ & $40.97 (31.62) \pm 1.11$ & $0.85 (45.4)$ & $\pm 1.15$ & $\pm 0.55$ & $\pm 1.69$\\
 $0.591$ & $26.003 (20.068) \pm 0.058$ & $40.72 (31.42) \pm 1.12$ & $0.88 (44.1)$ & $\pm 1.23$ & $\pm 0.36$ & $\pm 1.71$\\
 $0.569$ & $25.020 (19.289) \pm 0.056$ & $40.64 (31.33) \pm 1.15$ & $0.89 (42.6)$ & $\pm 1.43$ & $\pm 0.53$ & $\pm 1.91$\\
\end{tabular}
\end{ruledtabular}
\label{ap_06}
\end{table*}

\endgroup
\begingroup \squeezetable

\begin{table*}[tbh]
\caption{NBD fit results in centrality 30-40\%.
}
\begin{ruledtabular}
\begin{tabular}{ccccccc}
 $\delta\eta$ & $\langle \mu_{c} \rangle (\langle \mu \rangle)$ & $\langle k_{c} \rangle (\langle k \rangle)$ & $\langle \chi^{2}/NDF \rangle (\langle NDF \rangle)$ & $\delta \langle k_{c} \rangle (dead)$ & $\delta \langle k_{c} \rangle (fake)$ & $\delta \langle k_{c} \rangle (total)$ \\
\hline
 $0.700$ & $24.860 (19.241) \pm 0.058$ & $34.21 (26.48) \pm 0.88$ & $0.75 (47.0)$ & $\pm 1.14$ & $\pm 0.33$ & $\pm 1.47$\\
 $0.678$ & $24.177 (18.729) \pm 0.057$ & $34.00 (26.34) \pm 0.88$ & $0.77 (45.9)$ & $\pm 1.12$ & $\pm 0.30$ & $\pm 1.46$\\
 $0.656$ & $23.382 (18.088) \pm 0.055$ & $33.88 (26.21) \pm 0.90$ & $0.81 (44.2)$ & $\pm 0.91$ & $\pm 0.29$ & $\pm 1.31$\\
 $0.634$ & $22.573 (17.438) \pm 0.054$ & $33.73 (26.06) \pm 0.91$ & $0.90 (42.5)$ & $\pm 1.10$ & $\pm 0.25$ & $\pm 1.45$\\
 $0.613$ & $21.776 (16.808) \pm 0.053$ & $33.63 (25.96) \pm 0.93$ & $0.94 (40.9)$ & $\pm 1.08$ & $\pm 0.27$ & $\pm 1.44$\\
 $0.591$ & $20.986 (16.195) \pm 0.051$ & $33.54 (25.89) \pm 0.94$ & $0.92 (39.8)$ & $\pm 0.97$ & $\pm 0.25$ & $\pm 1.38$\\
 $0.569$ & $20.198 (15.571) \pm 0.050$ & $33.46 (25.80) \pm 0.97$ & $0.97 (38.4)$ & $\pm 1.04$ & $\pm 0.26$ & $\pm 1.44$\\
\end{tabular}
\end{ruledtabular}
\label{ap_07}
\end{table*}

\endgroup
\begingroup \squeezetable

\begin{table*}[tbh]
\caption{NBD fit results in centrality 35-45\%.
}
\begin{ruledtabular}
\begin{tabular}{ccccccc}
 $\delta\eta$ & $\langle \mu_{c} \rangle (\langle \mu \rangle)$ & $\langle k_{c} \rangle (\langle k \rangle)$ & $\langle \chi^{2}/NDF \rangle (\langle NDF \rangle)$ & $\delta \langle k_{c} \rangle (dead)$ & $\delta \langle k_{c} \rangle (fake)$ & $\delta \langle k_{c} \rangle (total)$ \\
\hline
 $0.700$ & $19.726 (15.268) \pm 0.051$ & $29.23 (22.62) \pm 0.80$ & $1.12 (40.0)$ & $\pm 0.69$ & $\pm 0.09$ & $\pm 1.06$\\
 $0.678$ & $19.191 (14.867) \pm 0.050$ & $29.24 (22.65) \pm 0.81$ & $1.02 (39.0)$ & $\pm 0.58$ & $\pm 0.17$ & $\pm 1.01$\\
 $0.656$ & $18.568 (14.364) \pm 0.049$ & $29.05 (22.47) \pm 0.82$ & $0.98 (38.2)$ & $\pm 0.76$ & $\pm 0.16$ & $\pm 1.13$\\
 $0.634$ & $17.920 (13.844) \pm 0.048$ & $29.04 (22.43) \pm 0.83$ & $1.09 (37.1)$ & $\pm 0.73$ & $\pm 0.13$ & $\pm 1.12$\\
 $0.613$ & $17.291 (13.348) \pm 0.046$ & $28.91 (22.31) \pm 0.85$ & $1.06 (36.2)$ & $\pm 0.77$ & $\pm 0.10$ & $\pm 1.15$\\
 $0.591$ & $16.672 (12.867) \pm 0.045$ & $28.64 (22.10) \pm 0.86$ & $1.04 (35.3)$ & $\pm 0.86$ & $\pm 0.20$ & $\pm 1.23$\\
 $0.569$ & $16.044 (12.369) \pm 0.044$ & $28.57 (22.02) \pm 0.88$ & $1.06 (34.2)$ & $\pm 0.79$ & $\pm 0.18$ & $\pm 1.20$\\
\end{tabular}
\end{ruledtabular}
\label{ap_08}
\end{table*}

\endgroup
\begingroup \squeezetable

\begin{table*}[tbh]
\caption{NBD fit results in centrality 40-50\%.
}
\begin{ruledtabular}
\begin{tabular}{ccccccc}
 $\delta\eta$ & $\langle \mu_{c} \rangle (\langle \mu \rangle)$ & $\langle k_{c} \rangle (\langle k \rangle)$ & $\langle \chi^{2}/NDF \rangle (\langle NDF \rangle)$ & $\delta \langle k_{c} \rangle (dead)$ & $\delta \langle k_{c} \rangle (fake)$ & $\delta \langle k_{c} \rangle (total)$ \\
\hline
 $0.700$ & $15.446 (11.955) \pm 0.044$ & $23.99 (18.57) \pm 0.65$ & $1.07 (36.0)$ & $\pm 1.02$ & $\pm 0.17$ & $\pm 1.22$\\
 $0.678$ & $15.032 (11.644) \pm 0.043$ & $23.87 (18.49) \pm 0.66$ & $1.08 (35.1)$ & $\pm 0.64$ & $\pm 0.18$ & $\pm 0.93$\\
 $0.656$ & $14.534 (11.243) \pm 0.042$ & $24.03 (18.59) \pm 0.68$ & $1.24 (34.0)$ & $\pm 0.61$ & $\pm 0.06$ & $\pm 0.91$\\
 $0.634$ & $14.033 (10.841) \pm 0.041$ & $24.11 (18.63) \pm 0.69$ & $1.25 (33.2)$ & $\pm 0.78$ & $\pm 0.16$ & $\pm 1.06$\\
 $0.613$ & $13.547 (10.457) \pm 0.041$ & $23.98 (18.51) \pm 0.71$ & $1.18 (32.4)$ & $\pm 0.77$ & $\pm 0.17$ & $\pm 1.06$\\
 $0.591$ & $13.055 (10.075) \pm 0.040$ & $23.97 (18.50) \pm 0.73$ & $1.15 (31.6)$ & $\pm 0.80$ & $\pm 0.16$ & $\pm 1.09$\\
 $0.569$ & $12.561 (9.684) \pm 0.038$ & $23.89 (18.42) \pm 0.74$ & $1.16 (30.8)$ & $\pm 0.68$ & $\pm 0.14$ & $\pm 1.02$\\
\end{tabular}
\end{ruledtabular}
\label{ap_09}
\end{table*}

\endgroup
\begingroup \squeezetable

\begin{table*}[tbh]
\caption{NBD fit results in centrality 45-55\%.
}
\begin{ruledtabular}
\begin{tabular}{ccccccc}
 $\delta\eta$ & $\langle \mu_{c} \rangle (\langle \mu \rangle)$ & $\langle k_{c} \rangle (\langle k \rangle)$ & $\langle \chi^{2}/NDF \rangle (\langle NDF \rangle)$ & $\delta \langle k_{c} \rangle (dead)$ & $\delta \langle k_{c} \rangle (fake)$ & $\delta \langle k_{c} \rangle (total)$ \\
\hline
 $0.700$ & $11.838 (9.163) \pm 0.038$ & $20.15 (15.59) \pm 0.61$ & $1.14 (28.0)$ & $\pm 0.71$ & $\pm 0.06$ & $\pm 0.94$\\
 $0.678$ & $11.517 (8.922) \pm 0.037$ & $20.06 (15.54) \pm 0.61$ & $1.10 (27.9)$ & $\pm 0.60$ & $\pm 0.03$ & $\pm 0.86$\\
 $0.656$ & $11.142 (8.620) \pm 0.036$ & $19.94 (15.43) \pm 0.62$ & $1.10 (26.6)$ & $\pm 0.48$ & $\pm 0.08$ & $\pm 0.79$\\
 $0.634$ & $10.759 (8.311) \pm 0.035$ & $19.94 (15.40) \pm 0.64$ & $1.08 (25.8)$ & $\pm 0.47$ & $\pm 0.11$ & $\pm 0.80$\\
 $0.613$ & $10.380 (8.013) \pm 0.035$ & $19.99 (15.43) \pm 0.65$ & $1.06 (24.9)$ & $\pm 0.51$ & $\pm 0.08$ & $\pm 0.83$\\
 $0.591$ & $10.005 (7.721) \pm 0.034$ & $19.95 (15.39) \pm 0.67$ & $1.06 (24.1)$ & $\pm 0.41$ & $\pm 0.11$ & $\pm 0.79$\\
 $0.569$ & $9.629 (7.424) \pm 0.033$ & $19.85 (15.30) \pm 0.68$ & $1.05 (23.2)$ & $\pm 0.44$ & $\pm 0.10$ & $\pm 0.82$\\
\end{tabular}
\end{ruledtabular}
\label{ap_10}
\end{table*}

\endgroup
\begingroup \squeezetable

\begin{table*}[tbh]
\caption{NBD fit results in centrality 50-60\%.
}
\begin{ruledtabular}
\begin{tabular}{ccccccc}
 $\delta\eta$ & $\langle \mu_{c} \rangle (\langle \mu \rangle)$ & $\langle k_{c} \rangle (\langle k \rangle)$ & $\langle \chi^{2}/NDF \rangle (\langle NDF \rangle)$ & $\delta \langle k_{c} \rangle (dead)$ & $\delta \langle k_{c} \rangle (fake)$ & $\delta \langle k_{c} \rangle (total)$ \\
\hline
 $0.700$ & $4.508 (3.489) \pm 0.023$ & $8.89 (6.88) \pm 0.29$ & $0.90 (15.0)$ & $\pm 0.21$ & $\pm 0.01$ & $\pm 0.35$\\
 $0.678$ & $4.387 (3.398) \pm 0.022$ & $8.85 (6.85) \pm 0.29$ & $0.90 (14.8)$ & $\pm 0.23$ & $\pm 0.01$ & $\pm 0.37$\\
 $0.656$ & $4.248 (3.286) \pm 0.022$ & $8.78 (6.79) \pm 0.29$ & $0.76 (14.2)$ & $\pm 0.25$ & $\pm 0.01$ & $\pm 0.39$\\
 $0.634$ & $4.102 (3.169) \pm 0.021$ & $8.74 (6.75) \pm 0.30$ & $0.76 (13.9)$ & $\pm 0.26$ & $\pm 0.01$ & $\pm 0.39$\\
 $0.613$ & $3.954 (3.052) \pm 0.021$ & $8.81 (6.80) \pm 0.31$ & $1.14 (13.4)$ & $\pm 0.23$ & $\pm 0.01$ & $\pm 0.38$\\
 $0.591$ & $3.813 (2.942) \pm 0.020$ & $8.67 (6.69) \pm 0.31$ & $0.91 (13.1)$ & $\pm 0.24$ & $\pm 0.01$ & $\pm 0.39$\\
 $0.569$ & $3.666 (2.827) \pm 0.020$ & $8.66 (6.67) \pm 0.32$ & $0.96 (12.9)$ & $\pm 0.27$ & $\pm 0.02$ & $\pm 0.42$\\
\end{tabular}
\end{ruledtabular}
\label{ap_11}
\end{table*}

\endgroup
\begingroup \squeezetable

\begin{table*}[tbh]
\caption{NBD fit results in centrality 55-65\%.
}
\begin{ruledtabular}
\begin{tabular}{ccccccc}
 $\delta\eta$ & $\langle \mu_{c} \rangle (\langle \mu \rangle)$ & $\langle k_{c} \rangle (\langle k \rangle)$ & $\langle \chi^{2}/NDF \rangle (\langle NDF \rangle)$ & $\delta \langle k_{c} \rangle (dead)$ & $\delta \langle k_{c} \rangle (fake)$ & $\delta \langle k_{c} \rangle (total)$ \\
\hline
 $0.700$ & $3.086 (2.389) \pm 0.019$ & $5.90 (4.56) \pm 0.20$ & $0.76 (13.0)$ & $\pm 0.25$ & $\pm 0.01$ & $\pm 0.32$\\
 $0.678$ & $3.000 (2.324) \pm 0.019$ & $5.88 (4.56) \pm 0.20$ & $0.79 (12.7)$ & $\pm 0.13$ & $\pm 0.01$ & $\pm 0.24$\\
 $0.656$ & $2.902 (2.245) \pm 0.018$ & $5.85 (4.52) \pm 0.20$ & $0.76 (12.5)$ & $\pm 0.14$ & $\pm 0.02$ & $\pm 0.24$\\
 $0.634$ & $2.803 (2.166) \pm 0.018$ & $5.78 (4.46) \pm 0.20$ & $0.69 (12.2)$ & $\pm 0.14$ & $\pm 0.02$ & $\pm 0.25$\\
 $0.613$ & $2.703 (2.086) \pm 0.017$ & $5.80 (4.48) \pm 0.21$ & $0.78 (12.1)$ & $\pm 0.07$ & $\pm 0.02$ & $\pm 0.22$\\
 $0.591$ & $2.607 (2.012) \pm 0.017$ & $5.76 (4.45) \pm 0.21$ & $0.70 (11.3)$ & $\pm 0.14$ & $\pm 0.02$ & $\pm 0.25$\\
 $0.569$ & $2.508 (1.934) \pm 0.017$ & $5.76 (4.44) \pm 0.22$ & $0.62 (10.9)$ & $\pm 0.13$ & $\pm 0.02$ & $\pm 0.25$\\
\end{tabular}
\end{ruledtabular}
\label{ap_12}
\end{table*}

\endgroup
\begingroup \squeezetable

\begin{table*}[tbh]
\caption{NBD fit results in centrality 0-5\%.
}
\begin{ruledtabular}
\begin{tabular}{ccccccc}
 $\delta\eta$ & $\langle \mu_{c} \rangle (\langle \mu \rangle)$ & $\langle k_{c} \rangle (\langle k \rangle)$ & $\langle \chi^{2}/NDF \rangle (\langle NDF \rangle)$ & $\delta \langle k_{c} \rangle (dead)$ & $\delta \langle k_{c} \rangle (fake)$ & $\delta \langle k_{c} \rangle (total)$ \\
\hline
 $0.700$ & $84.371 (65.303) \pm 0.148$ & $341.22 (264.10) \pm 31.75$ & $0.54 (58.0)$ & $\pm 54.10$ & $\pm 6.72$ & $\pm 63.09$\\
 $0.678$ & $82.149 (63.638) \pm 0.144$ & $359.73 (278.66) \pm 35.51$ & $0.58 (57.8)$ & $\pm 29.67$ & $\pm 26.46$ & $\pm 53.31$\\
 $0.656$ & $79.451 (61.463) \pm 0.141$ & $369.96 (286.25) \pm 37.82$ & $0.66 (56.4)$ & $\pm 13.85$ & $\pm 14.18$ & $\pm 42.69$\\
 $0.634$ & $76.740 (59.282) \pm 0.138$ & $370.46 (286.18) \pm 38.62$ & $0.63 (55.4)$ & $\pm 19.09$ & $\pm 14.98$ & $\pm 45.61$\\
 $0.613$ & $74.027 (57.142) \pm 0.135$ & $372.37 (287.35) \pm 40.79$ & $0.62 (54.7)$ & $\pm 13.54$ & $\pm 15.22$ & $\pm 45.60$\\
 $0.591$ & $71.331 (55.050) \pm 0.133$ & $374.07 (288.55) \pm 42.84$ & $0.64 (53.6)$ & $\pm 13.41$ & $\pm 17.15$ & $\pm 48.06$\\
 $0.569$ & $68.629 (52.910) \pm 0.130$ & $374.10 (288.36) \pm 44.31$ & $0.64 (52.7)$ & $\pm 18.44$ & $\pm 19.49$ & $\pm 51.80$\\
\end{tabular}
\end{ruledtabular}
\label{ap_13}
\end{table*}

\endgroup
\begingroup \squeezetable

\begin{table*}[tbh]
\caption{NBD fit results in centrality 5-10\%.
}
\begin{ruledtabular}
\begin{tabular}{ccccccc}
 $\delta\eta$ & $\langle \mu_{c} \rangle (\langle \mu \rangle)$ & $\langle k_{c} \rangle (\langle k \rangle)$ & $\langle \chi^{2}/NDF \rangle (\langle NDF \rangle)$ & $\delta \langle k_{c} \rangle (dead)$ & $\delta \langle k_{c} \rangle (fake)$ & $\delta \langle k_{c} \rangle (total)$ \\
\hline
 $0.700$ & $72.246 (55.918) \pm 0.126$ & $187.62 (145.22) \pm 10.56$ & $0.80 (61.0)$ & $\pm 1.88$ & $\pm 9.61$ & $\pm 14.40$\\
 $0.678$ & $70.291 (54.452) \pm 0.124$ & $185.02 (143.32) \pm 10.61$ & $0.68 (61.3)$ & $\pm 10.46$ & $\pm 6.69$ & $\pm 16.33$\\
 $0.656$ & $67.996 (52.602) \pm 0.121$ & $188.09 (145.50) \pm 11.00$ & $0.71 (59.8)$ & $\pm 4.99$ & $\pm 4.56$ & $\pm 12.91$\\
 $0.634$ & $65.656 (50.720) \pm 0.119$ & $188.72 (145.76) \pm 11.29$ & $0.76 (59.0)$ & $\pm 2.81$ & $\pm 6.48$ & $\pm 13.32$\\
 $0.613$ & $63.346 (48.897) \pm 0.116$ & $186.50 (143.95) \pm 11.39$ & $0.79 (57.0)$ & $\pm 4.55$ & $\pm 6.31$ & $\pm 13.79$\\
 $0.591$ & $61.055 (47.118) \pm 0.114$ & $185.01 (142.75) \pm 11.53$ & $0.76 (55.6)$ & $\pm 6.00$ & $\pm 5.32$ & $\pm 14.04$\\
 $0.569$ & $58.752 (45.296) \pm 0.111$ & $184.61 (142.29) \pm 11.90$ & $0.76 (54.0)$ & $\pm 4.96$ & $\pm 7.98$ & $\pm 15.16$\\
\end{tabular}
\end{ruledtabular}
\label{ap_14}
\end{table*}

\endgroup
\begingroup \squeezetable

\begin{table*}[tbh]
\caption{NBD fit results in centrality 10-15\%.
}
\begin{ruledtabular}
\begin{tabular}{ccccccc}
 $\delta\eta$ & $\langle \mu_{c} \rangle (\langle \mu \rangle)$ & $\langle k_{c} \rangle (\langle k \rangle)$ & $\langle \chi^{2}/NDF \rangle (\langle NDF \rangle)$ & $\delta \langle k_{c} \rangle (dead)$ & $\delta \langle k_{c} \rangle (fake)$ & $\delta \langle k_{c} \rangle (total)$ \\
\hline
 $0.700$ & $60.607 (46.909) \pm 0.119$ & $129.59 (100.31) \pm 6.29$ & $0.80 (58.0)$ & $\pm 1.29$ & $\pm 1.51$ & $\pm 6.60$\\
 $0.678$ & $59.026 (45.725) \pm 0.118$ & $123.34 (95.54) \pm 6.09$ & $0.76 (56.6)$ & $\pm 4.71$ & $\pm 9.77$ & $\pm 12.44$\\
 $0.656$ & $57.108 (44.179) \pm 0.116$ & $121.44 (93.94) \pm 6.08$ & $0.70 (55.3)$ & $\pm 5.55$ & $\pm 4.90$ & $\pm 9.58$\\
 $0.634$ & $55.139 (42.596) \pm 0.113$ & $121.88 (94.14) \pm 6.24$ & $0.70 (54.2)$ & $\pm 4.92$ & $\pm 3.01$ & $\pm 8.50$\\
 $0.613$ & $53.198 (41.064) \pm 0.111$ & $120.98 (93.37) \pm 6.33$ & $0.69 (52.9)$ & $\pm 5.23$ & $\pm 2.63$ & $\pm 8.62$\\
 $0.591$ & $51.256 (39.556) \pm 0.108$ & $120.49 (92.96) \pm 6.46$ & $0.71 (52.2)$ & $\pm 3.80$ & $\pm 3.87$ & $\pm 8.44$\\
 $0.569$ & $49.321 (38.025) \pm 0.106$ & $119.78 (92.31) \pm 6.58$ & $0.70 (51.3)$ & $\pm 3.61$ & $\pm 2.65$ & $\pm 7.95$\\
\end{tabular}
\end{ruledtabular}
\label{ap_15}
\end{table*}

\endgroup
\begingroup \squeezetable

\begin{table*}[tbh]
\caption{NBD fit results in centrality 15-20\%.
}
\begin{ruledtabular}
\begin{tabular}{ccccccc}
 $\delta\eta$ & $\langle \mu_{c} \rangle (\langle \mu \rangle)$ & $\langle k_{c} \rangle (\langle k \rangle)$ & $\langle \chi^{2}/NDF \rangle (\langle NDF \rangle)$ & $\delta \langle k_{c} \rangle (dead)$ & $\delta \langle k_{c} \rangle (fake)$ & $\delta \langle k_{c} \rangle (total)$ \\
\hline
 $0.700$ & $50.243 (38.888) \pm 0.109$ & $95.08 (73.59) \pm 4.42$ & $0.79 (53.0)$ & $\pm 4.84$ & $\pm 6.50$ & $\pm 9.23$\\
 $0.678$ & $48.892 (37.875) \pm 0.106$ & $95.76 (74.18) \pm 4.54$ & $0.67 (52.4)$ & $\pm 3.20$ & $\pm 1.41$ & $\pm 5.73$\\
 $0.656$ & $47.301 (36.592) \pm 0.104$ & $95.63 (73.98) \pm 4.65$ & $0.75 (51.2)$ & $\pm 2.80$ & $\pm 1.66$ & $\pm 5.68$\\
 $0.634$ & $45.685 (35.292) \pm 0.102$ & $95.70 (73.94) \pm 4.78$ & $0.71 (49.2)$ & $\pm 2.09$ & $\pm 1.96$ & $\pm 5.57$\\
 $0.613$ & $44.087 (34.031) \pm 0.100$ & $94.78 (73.17) \pm 4.87$ & $0.71 (47.7)$ & $\pm 3.79$ & $\pm 3.17$ & $\pm 6.94$\\
 $0.591$ & $42.483 (32.785) \pm 0.097$ & $95.97 (74.07) \pm 5.05$ & $0.70 (47.1)$ & $\pm 2.95$ & $\pm 1.99$ & $\pm 6.18$\\
 $0.569$ & $40.884 (31.519) \pm 0.095$ & $96.10 (74.10) \pm 5.22$ & $0.69 (46.1)$ & $\pm 2.66$ & $\pm 2.10$ & $\pm 6.22$\\
\end{tabular}
\end{ruledtabular}
\label{ap_16}
\end{table*}

\endgroup
\begingroup \squeezetable

\begin{table*}[tbh]
\caption{NBD fit results in centrality 20-25\%.
}
\begin{ruledtabular}
\begin{tabular}{ccccccc}
 $\delta\eta$ & $\langle \mu_{c} \rangle (\langle \mu \rangle)$ & $\langle k_{c} \rangle (\langle k \rangle)$ & $\langle \chi^{2}/NDF \rangle (\langle NDF \rangle)$ & $\delta \langle k_{c} \rangle (dead)$ & $\delta \langle k_{c} \rangle (fake)$ & $\delta \langle k_{c} \rangle (total)$ \\
\hline
 $0.700$ & $41.340 (31.997) \pm 0.101$ & $80.17 (62.05) \pm 3.85$ & $0.71 (48.0)$ & $\pm 1.34$ & $\pm 1.27$ & $\pm 4.27$\\
 $0.678$ & $40.233 (31.167) \pm 0.099$ & $80.14 (62.08) \pm 3.92$ & $0.67 (47.2)$ & $\pm 3.19$ & $\pm 1.69$ & $\pm 5.33$\\
 $0.656$ & $38.920 (30.109) \pm 0.096$ & $81.43 (63.00) \pm 4.07$ & $0.68 (46.9)$ & $\pm 1.52$ & $\pm 0.92$ & $\pm 4.44$\\
 $0.634$ & $37.606 (29.051) \pm 0.094$ & $81.18 (62.71) \pm 4.17$ & $0.68 (45.8)$ & $\pm 3.68$ & $\pm 1.51$ & $\pm 5.76$\\
 $0.613$ & $36.282 (28.006) \pm 0.092$ & $82.48 (63.67) \pm 4.38$ & $0.78 (44.7)$ & $\pm 1.90$ & $\pm 1.68$ & $\pm 5.07$\\
 $0.591$ & $34.986 (27.000) \pm 0.090$ & $81.78 (63.12) \pm 4.43$ & $0.76 (44.2)$ & $\pm 2.76$ & $\pm 1.37$ & $\pm 5.39$\\
 $0.569$ & $33.656 (25.947) \pm 0.088$ & $81.76 (63.04) \pm 4.53$ & $0.75 (43.4)$ & $\pm 1.38$ & $\pm 1.07$ & $\pm 4.86$\\
\end{tabular}
\end{ruledtabular}
\label{ap_17}
\end{table*}

\endgroup
\begingroup \squeezetable

\begin{table*}[tbh]
\caption{NBD fit results in centrality 25-30\%.
}
\begin{ruledtabular}
\begin{tabular}{ccccccc}
 $\delta\eta$ & $\langle \mu_{c} \rangle (\langle \mu \rangle)$ & $\langle k_{c} \rangle (\langle k \rangle)$ & $\langle \chi^{2}/NDF \rangle (\langle NDF \rangle)$ & $\delta \langle k_{c} \rangle (dead)$ & $\delta \langle k_{c} \rangle (fake)$ & $\delta \langle k_{c} \rangle (total)$ \\
\hline
 $0.700$ & $33.991 (26.309) \pm 0.090$ & $67.99 (52.63) \pm 3.18$ & $0.56 (44.0)$ & $\pm 2.58$ & $\pm 0.80$ & $\pm 4.17$\\
 $0.678$ & $33.067 (25.615) \pm 0.088$ & $68.77 (53.27) \pm 3.29$ & $0.61 (44.0)$ & $\pm 2.56$ & $\pm 0.05$ & $\pm 4.17$\\
 $0.656$ & $31.992 (24.750) \pm 0.086$ & $67.53 (52.23) \pm 3.30$ & $0.56 (43.6)$ & $\pm 3.20$ & $\pm 1.31$ & $\pm 4.78$\\
 $0.634$ & $30.890 (23.863) \pm 0.084$ & $67.26 (51.95) \pm 3.34$ & $0.58 (42.8)$ & $\pm 2.31$ & $\pm 1.01$ & $\pm 4.19$\\
 $0.613$ & $29.811 (23.012) \pm 0.082$ & $67.40 (52.02) \pm 3.41$ & $0.64 (41.9)$ & $\pm 1.64$ & $\pm 1.27$ & $\pm 3.99$\\
 $0.591$ & $28.732 (22.173) \pm 0.081$ & $66.18 (51.07) \pm 3.41$ & $0.65 (41.3)$ & $\pm 2.18$ & $\pm 0.79$ & $\pm 4.12$\\
 $0.569$ & $27.643 (21.312) \pm 0.079$ & $66.17 (51.01) \pm 3.50$ & $0.62 (40.2)$ & $\pm 2.87$ & $\pm 1.24$ & $\pm 4.69$\\
\end{tabular}
\end{ruledtabular}
\label{ap_18}
\end{table*}

\clearpage

\endgroup
\begingroup \squeezetable

\begin{table*}[tbh]
\caption{NBD fit results in centrality 30-35\%.
}
\begin{ruledtabular}
\begin{tabular}{ccccccc}
 $\delta\eta$ & $\langle \mu_{c} \rangle (\langle \mu \rangle)$ & $\langle k_{c} \rangle (\langle k \rangle)$ & $\langle \chi^{2}/NDF \rangle (\langle NDF \rangle)$ & $\delta \langle k_{c} \rangle (dead)$ & $\delta \langle k_{c} \rangle (fake)$ & $\delta \langle k_{c} \rangle (total)$ \\
\hline
 $0.700$ & $27.652 (21.402) \pm 0.079$ & $58.33 (45.15) \pm 2.86$ & $0.67 (41.0)$ & $\pm 3.24$ & $\pm 0.86$ & $\pm 4.41$\\
 $0.678$ & $26.900 (20.838) \pm 0.078$ & $57.89 (44.85) \pm 2.86$ & $0.60 (40.9)$ & $\pm 2.95$ & $\pm 0.85$ & $\pm 4.20$\\
 $0.656$ & $26.016 (20.126) \pm 0.076$ & $57.99 (44.86) \pm 2.94$ & $0.63 (39.6)$ & $\pm 2.17$ & $\pm 0.88$ & $\pm 3.75$\\
 $0.634$ & $25.115 (19.401) \pm 0.074$ & $57.51 (44.43) \pm 2.98$ & $0.66 (38.4)$ & $\pm 2.60$ & $\pm 0.72$ & $\pm 4.02$\\
 $0.613$ & $24.225 (18.699) \pm 0.073$ & $56.95 (43.97) \pm 3.03$ & $0.67 (37.0)$ & $\pm 2.60$ & $\pm 0.86$ & $\pm 4.09$\\
 $0.591$ & $23.345 (18.015) \pm 0.071$ & $56.94 (43.95) \pm 3.10$ & $0.70 (36.3)$ & $\pm 2.27$ & $\pm 0.77$ & $\pm 3.92$\\
 $0.569$ & $22.464 (17.318) \pm 0.070$ & $56.67 (43.71) \pm 3.17$ & $0.74 (35.0)$ & $\pm 1.78$ & $\pm 0.77$ & $\pm 3.72$\\
\end{tabular}
\end{ruledtabular}
\label{ap_19}
\end{table*}

\endgroup
\begingroup \squeezetable

\begin{table*}[tbh]
\caption{NBD fit results in centrality 35-40\%.
}
\begin{ruledtabular}
\begin{tabular}{ccccccc}
 $\delta\eta$ & $\langle \mu_{c} \rangle (\langle \mu \rangle)$ & $\langle k_{c} \rangle (\langle k \rangle)$ & $\langle \chi^{2}/NDF \rangle (\langle NDF \rangle)$ & $\delta \langle k_{c} \rangle (dead)$ & $\delta \langle k_{c} \rangle (fake)$ & $\delta \langle k_{c} \rangle (total)$ \\
\hline
 $0.700$ & $21.970 (17.005) \pm 0.071$ & $47.25 (36.57) \pm 2.31$ & $0.79 (38.0)$ & $\pm 0.68$ & $\pm 0.60$ & $\pm 2.48$\\
 $0.678$ & $21.372 (16.556) \pm 0.070$ & $47.47 (36.77) \pm 2.32$ & $0.69 (37.7)$ & $\pm 0.44$ & $\pm 0.02$ & $\pm 2.36$\\
 $0.656$ & $20.684 (16.001) \pm 0.068$ & $46.97 (36.33) \pm 2.33$ & $0.71 (37.1)$ & $\pm 0.92$ & $\pm 0.23$ & $\pm 2.51$\\
 $0.634$ & $19.964 (15.423) \pm 0.067$ & $46.50 (35.92) \pm 2.35$ & $0.79 (36.0)$ & $\pm 0.86$ & $\pm 0.30$ & $\pm 2.52$\\
 $0.613$ & $19.253 (14.862) \pm 0.065$ & $46.36 (35.78) \pm 2.40$ & $0.86 (34.6)$ & $\pm 0.91$ & $\pm 0.15$ & $\pm 2.57$\\
 $0.591$ & $18.572 (14.333) \pm 0.064$ & $45.46 (35.08) \pm 2.41$ & $0.82 (33.4)$ & $\pm 1.16$ & $\pm 0.36$ & $\pm 2.70$\\
 $0.569$ & $17.877 (13.783) \pm 0.063$ & $45.55 (35.11) \pm 2.48$ & $0.83 (32.5)$ & $\pm 1.22$ & $\pm 0.37$ & $\pm 2.79$\\
\end{tabular}
\end{ruledtabular}
\label{ap_20}
\end{table*}

\endgroup
\begingroup \squeezetable

\begin{table*}[tbh]
\caption{NBD fit results in centrality 40-45\%.
}
\begin{ruledtabular}
\begin{tabular}{ccccccc}
 $\delta\eta$ & $\langle \mu_{c} \rangle (\langle \mu \rangle)$ & $\langle k_{c} \rangle (\langle k \rangle)$ & $\langle \chi^{2}/NDF \rangle (\langle NDF \rangle)$ & $\delta \langle k_{c} \rangle (dead)$ & $\delta \langle k_{c} \rangle (fake)$ & $\delta \langle k_{c} \rangle (total)$ \\
\hline
 $0.700$ & $17.453 (13.508) \pm 0.063$ & $37.64 (29.13) \pm 1.83$ & $0.74 (35.0)$ & $\pm 0.40$ & $\pm 0.98$ & $\pm 2.12$\\
 $0.678$ & $16.988 (13.160) \pm 0.062$ & $37.36 (28.94) \pm 1.85$ & $0.74 (34.1)$ & $\pm 1.08$ & $\pm 0.27$ & $\pm 2.16$\\
 $0.656$ & $16.429 (12.709) \pm 0.061$ & $37.57 (29.06) \pm 1.91$ & $0.83 (33.0)$ & $\pm 0.91$ & $\pm 0.00$ & $\pm 2.12$\\
 $0.634$ & $15.863 (12.255) \pm 0.059$ & $37.55 (29.01) \pm 1.96$ & $0.82 (32.2)$ & $\pm 1.54$ & $\pm 0.31$ & $\pm 2.51$\\
 $0.613$ & $15.315 (11.822) \pm 0.058$ & $36.88 (28.47) \pm 1.96$ & $0.76 (31.4)$ & $\pm 1.75$ & $\pm 0.34$ & $\pm 2.65$\\
 $0.591$ & $14.758 (11.389) \pm 0.057$ & $36.84 (28.43) \pm 2.00$ & $0.83 (30.6)$ & $\pm 1.76$ & $\pm 0.32$ & $\pm 2.68$\\
 $0.569$ & $14.201 (10.948) \pm 0.056$ & $36.74 (28.32) \pm 2.04$ & $0.88 (29.8)$ & $\pm 1.32$ & $\pm 0.26$ & $\pm 2.45$\\
\end{tabular}
\end{ruledtabular}
\label{ap_21}
\end{table*}

\endgroup
\begingroup \squeezetable

\begin{table*}[tbh]
\caption{NBD fit results in centrality 45-50\%.
}
\begin{ruledtabular}
\begin{tabular}{ccccccc}
 $\delta\eta$ & $\langle \mu_{c} \rangle (\langle \mu \rangle)$ & $\langle k_{c} \rangle (\langle k \rangle)$ & $\langle \chi^{2}/NDF \rangle (\langle NDF \rangle)$ & $\delta \langle k_{c} \rangle (dead)$ & $\delta \langle k_{c} \rangle (fake)$ & $\delta \langle k_{c} \rangle (total)$ \\
\hline
 $0.700$ & $13.493 (10.444) \pm 0.054$ & $32.22 (24.94) \pm 1.72$ & $0.77 (27.0)$ & $\pm 1.57$ & $\pm 0.13$ & $\pm 2.34$\\
 $0.678$ & $13.123 (10.166) \pm 0.053$ & $32.44 (25.13) \pm 1.74$ & $0.79 (26.9)$ & $\pm 1.12$ & $\pm 0.09$ & $\pm 2.07$\\
 $0.656$ & $12.698 (9.823) \pm 0.052$ & $32.30 (24.99) \pm 1.77$ & $0.81 (25.6)$ & $\pm 0.70$ & $\pm 0.22$ & $\pm 1.92$\\
 $0.634$ & $12.261 (9.472) \pm 0.051$ & $32.54 (25.14) \pm 1.85$ & $0.89 (24.6)$ & $\pm 0.78$ & $\pm 0.32$ & $\pm 2.04$\\
 $0.613$ & $11.829 (9.131) \pm 0.050$ & $32.98 (25.46) \pm 1.94$ & $0.89 (23.9)$ & $\pm 0.81$ & $\pm 0.20$ & $\pm 2.11$\\
 $0.591$ & $11.403 (8.800) \pm 0.049$ & $33.01 (25.47) \pm 2.00$ & $0.88 (23.1)$ & $\pm 0.56$ & $\pm 0.28$ & $\pm 2.10$\\
 $0.569$ & $10.976 (8.462) \pm 0.048$ & $32.73 (25.23) \pm 2.04$ & $0.84 (22.3)$ & $\pm 0.56$ & $\pm 0.24$ & $\pm 2.13$\\
\end{tabular}
\end{ruledtabular}
\label{ap_22}
\end{table*}

\endgroup
\begingroup \squeezetable

\begin{table*}[tbh]
\caption{NBD fit results in centrality 50-55\%.
}
\begin{ruledtabular}
\begin{tabular}{ccccccc}
 $\delta\eta$ & $\langle \mu_{c} \rangle (\langle \mu \rangle)$ & $\langle k_{c} \rangle (\langle k \rangle)$ & $\langle \chi^{2}/NDF \rangle (\langle NDF \rangle)$ & $\delta \langle k_{c} \rangle (dead)$ & $\delta \langle k_{c} \rangle (fake)$ & $\delta \langle k_{c} \rangle (total)$ \\
\hline
 $0.700$ & $10.177 (7.877) \pm 0.046$ & $27.08 (20.96) \pm 1.52$ & $1.18 (24.0)$ & $\pm 1.23$ & $\pm 0.17$ & $\pm 1.96$\\
 $0.678$ & $9.898 (7.667) \pm 0.045$ & $26.98 (20.90) \pm 1.53$ & $1.14 (23.0)$ & $\pm 0.62$ & $\pm 0.26$ & $\pm 1.67$\\
 $0.656$ & $9.584 (7.414) \pm 0.044$ & $26.02 (20.12) \pm 1.50$ & $1.07 (21.9)$ & $\pm 0.94$ & $\pm 0.23$ & $\pm 1.79$\\
 $0.634$ & $9.257 (7.151) \pm 0.044$ & $25.65 (19.81) \pm 1.51$ & $0.97 (21.2)$ & $\pm 0.43$ & $\pm 0.21$ & $\pm 1.58$\\
 $0.613$ & $8.935 (6.897) \pm 0.043$ & $25.16 (19.42) \pm 1.52$ & $0.85 (20.8)$ & $\pm 0.63$ & $\pm 0.16$ & $\pm 1.66$\\
 $0.591$ & $8.613 (6.647) \pm 0.042$ & $24.75 (19.10) \pm 1.53$ & $0.82 (20.0)$ & $\pm 0.77$ & $\pm 0.21$ & $\pm 1.73$\\
 $0.569$ & $8.287 (6.389) \pm 0.041$ & $24.69 (19.03) \pm 1.58$ & $0.80 (19.6)$ & $\pm 0.54$ & $\pm 0.10$ & $\pm 1.68$\\
\end{tabular}
\end{ruledtabular}
\label{ap_23}
\end{table*}

\endgroup
\begingroup \squeezetable

\begin{table*}[h]
\caption{NBD fit results in centrality 55-60\%.}
\begin{ruledtabular}
\begin{tabular}{ccccccc}
 $\delta\eta$ & $\langle \mu_{c} \rangle (\langle \mu \rangle)$ & $\langle k_{c} \rangle (\langle k \rangle)$ & $\langle \chi^{2}/NDF \rangle (\langle NDF \rangle)$ & $\delta \langle k_{c} \rangle (dead)$ & $\delta \langle k_{c} \rangle (fake)$ & $\delta \langle k_{c} \rangle (total)$ \\
\hline
 $0.700$ & $7.465 (5.778) \pm 0.041$ & $18.75 (14.52) \pm 1.12$ & $0.71 (16.0)$ & $\pm 0.59$ & $\pm 0.13$ & $\pm 1.27$\\
 $0.678$ & $7.258 (5.622) \pm 0.040$ & $19.02 (14.74) \pm 1.14$ & $0.76 (16.0)$ & $\pm 0.60$ & $\pm 0.08$ & $\pm 1.29$\\
 $0.656$ & $7.021 (5.431) \pm 0.039$ & $18.80 (14.54) \pm 1.14$ & $0.73 (15.8)$ & $\pm 0.61$ & $\pm 0.09$ & $\pm 1.29$\\
 $0.634$ & $6.781 (5.238) \pm 0.038$ & $18.41 (14.22) \pm 1.14$ & $0.57 (15.4)$ & $\pm 0.57$ & $\pm 0.09$ & $\pm 1.28$\\
 $0.613$ & $6.547 (5.054) \pm 0.038$ & $18.01 (13.89) \pm 1.13$ & $0.51 (14.7)$ & $\pm 0.70$ & $\pm 0.07$ & $\pm 1.33$\\
 $0.591$ & $6.307 (4.867) \pm 0.037$ & $17.98 (13.87) \pm 1.15$ & $0.56 (14.4)$ & $\pm 0.70$ & $\pm 0.07$ & $\pm 1.34$\\
 $0.569$ & $6.070 (4.680) \pm 0.036$ & $17.84 (13.75) \pm 1.15$ & $0.51 (14.1)$ & $\pm 0.69$ & $\pm 0.07$ & $\pm 1.35$\\
\end{tabular}
\end{ruledtabular}
\label{ap_24}
\end{table*}

\endgroup


\begingroup \squeezetable
\begin{table*}[tbh]
\caption{NBD fit results in centrality 60-65\%.}
\begin{ruledtabular}
\begin{tabular}{ccccccc}
 $\delta\eta$ & $\langle \mu_{c} \rangle (\langle \mu \rangle)$ & $\langle k_{c} \rangle (\langle k \rangle)$ & $\langle \chi^{2}/NDF \rangle (\langle NDF \rangle)$ & $\delta \langle k_{c} \rangle (dead)$ & $\delta \langle k_{c} \rangle (fake)$ & $\delta \langle k_{c} \rangle (total)$ \\
\hline
 $0.700$ & $5.337 (4.131) \pm 0.033$ & $14.40 (11.14) \pm 0.81$ & $0.79 (15.0)$ & $\pm 0.33$ & $\pm 0.04$ & $\pm 0.88$\\
 $0.678$ & $5.197 (4.026) \pm 0.033$ & $14.25 (11.04) \pm 0.82$ & $0.77 (14.8)$ & $\pm 0.45$ & $\pm 0.04$ & $\pm 0.94$\\
 $0.656$ & $5.031 (3.892) \pm 0.032$ & $14.37 (11.12) \pm 0.86$ & $0.61 (14.2)$ & $\pm 0.46$ & $\pm 0.03$ & $\pm 0.97$\\
 $0.634$ & $4.861 (3.755) \pm 0.032$ & $14.35 (11.09) \pm 0.88$ & $0.53 (13.8)$ & $\pm 0.47$ & $\pm 0.04$ & $\pm 0.99$\\
 $0.613$ & $4.686 (3.617) \pm 0.031$ & $14.41 (11.13) \pm 0.93$ & $0.78 (13.1)$ & $\pm 0.44$ & $\pm 0.04$ & $\pm 1.03$\\
 $0.591$ & $4.519 (3.487) \pm 0.030$ & $13.99 (10.80) \pm 0.90$ & $0.63 (12.9)$ & $\pm 0.53$ & $\pm 0.04$ & $\pm 1.05$\\
 $0.569$ & $4.344 (3.349) \pm 0.029$ & $14.01 (10.80) \pm 0.92$ & $0.73 (12.7)$ & $\pm 0.51$ & $\pm 0.05$ & $\pm 1.06$\\
\end{tabular}
\end{ruledtabular}
\label{ap_25}
\end{table*}
\endgroup

\end{appendix}

\clearpage


\end{document}